%% file: tt3j.tex
\documentclass[twocolumn,epjc3]{svjour3}
\pdfoutput=1
\usepackage[T1]{fontenc}
\usepackage{amsmath,amssymb,pstricks,graphicx,xspace,subfigure}
\usepackage{flushend}
\usepackage{cite}
\usepackage[colorlinks,citecolor=blue,urlcolor=blue,linkcolor=blue]{hyperref}
\journalname{Eur. Phys. J. C}
\newcommand{\Sherpa}{S\protect\scalebox{0.8}{HERPA}\xspace}
\newcommand{\OpenLoops}{O\protect\scalebox{0.8}{PEN}L\protect\scalebox{0.8}{OOPS}\xspace}
\newcommand{\Comix}{C\protect\scalebox{0.8}{OMIX}\xspace}

\newcommand{\Collier}{C\protect\scalebox{0.8}{OLLIER}\xspace}
\newcommand{\CutTools}{C\protect\scalebox{0.8}{UT}T\protect\scalebox{0.8}{OOLS}\xspace}
\newcommand{\OneLOop}{O\protect\scalebox{0.8}{NE}LO\protect\scalebox{0.8}{OP}\xspace}
\newcommand{\Rivet}{R\protect\scalebox{0.8}{IVET}\xspace}
\newcommand{\MINLO}{M\protect\scalebox{0.8}{I}NLO\xspace}

\def\rT{\mathrm{T}}
\def\pT{p_{\mathrm{T}}}
\def\jet{\mathrm{jet}}

\def\ttbar{t\bar t}
\def\capskip{\vskip -2mm}
\def\HT{H_\rT}
\def\htjets{H_\rT^{\mathrm{jets}}}
\def\ptjet{p_{\rT,\mathrm{jet}}}
\newcommand{\hskipthree}{\hskip 2pt}
\newcommand{\hskipfour}{\hskip 1pt}
\def\njets{N_{\mathrm{jets}}}
\def\ttbar{t\bar t}
\newcommand{\reffi}[1]{Fig.~\ref{#1}}

\def\as{\alpha_{\mathrm{s}}}
\def\mucore{\mu_{\mathrm{core}}}
\def\muR{\mu_{\mathrm{R}}}
\def\muF{\mu_{\mathrm{F}}}
\def\xiF{\xi_{\mathrm{F}}}
\def\qmin{q_{\mathrm{min}}}
\def\abtilde{\tilde{a}\tilde{b}}
\def\tN{\tilde{N}}

\begin{document}

\title{Next-to-leading order QCD predictions for top-quark pair production with up to three jets}


\author{
   S.~H{\"o}che\thanksref{e1,addr1}
   \and P.~Maierh\"ofer\thanksref{e2,addr2} 
   \and N.~Moretti\thanksref{e3,addr3} 
   \and S.~Pozzorini\thanksref{e4,addr3}
   \and F.~Siegert\thanksref{e5,addr4}}

\thankstext{e1}{e-mail: shoeche@slac.stanford.edu}
\thankstext{e2}{e-mail: philipp.maierhoefer@physik.uni-freiburg.de}
\thankstext{e3}{e-mail: moretti@physik.uzh.ch}
\thankstext{e4}{e-mail: pozzorin@physik.uzh.ch}
\thankstext{e5}{e-mail: frank.siegert@cern.ch}


\institute{SLAC National Accelerator Laboratory, Menlo Park, CA 94025, USA \label{addr1}
   \and Physikalisches Institut, Albert-Ludwigs-Universit{\"a}t Freiburg, D-79104 Freiburg, Germany\label{addr2}
   \and Physik--Institut, Universit{\"a}t Z{\"u}rich, CH--8057 Z{\"u}rich, Switzerland\label{addr3}
   \and Institut f{\"u}r Kern- und Teilchenphysik, TU Dresden, D--01062 Dresden, Germany\label{addr4}
\vspace{19mm}
}

\date{}

\maketitle

\begin{abstract}

We present theoretical predictions for
the production of top-quark pairs with up to three jets at the next-to leading
order in perturbative QCD. The relevant calculations are performed 
with \Sherpa and \OpenLoops. To address the issue of scale choices and 
related uncertainties 
in the presence of multiple scales, we compare results obtained 
with the standard scale $\HT/2$ at fixed order and the \MINLO procedure.  
Analyzing various cross sections and distributions for $\ttbar+0,1,2,3$\,jets at the 13\,TeV
LHC we find a remarkable overall agreement between fixed-order and \MINLO results.
The differences are typically below the respective factor-two scale
variations, suggesting that for all considered jet multiplicities
missing higher-order effects should not exceed the
ten percent level.
\PACS{
12.38.--t,  
12.38.Bx,   
13.85.--t,  
14.65.Ha    
}

\end{abstract}


\section{Introduction}

The top quark as the heaviest known elementary particle plays a
fundamental role, both in the Standard Model and in new physics scenarios.
Experimental analyses of Large Hadron Collider (LHC) data collected during
run II will provide unprecedented reach at high energy and in exclusive
phase space regions with associated production of jets and vector bosons or Higgs bosons. 
The production of a $\ttbar$ system in association with multiple jets plays an
especially important role as a background to new physics searches and 
to various Higgs and Standard Model analyses.  In particular, the precise
theoretical control of $\ttbar+$multijet backgrounds is one of the most
important prerequisites for the observation of top-quark production in
association with a Higgs boson, which would give direct access to the
top-quark Yukawa coupling.
In addition, $\ttbar+$multijet production allows for
powerful test of perturbative QCD and is also routinely exploited 
for the validation of Monte Carlo tools that are used 
in a multitude of LHC studies.
All these analyses require theoretical predictions at the highest possible accuracy.  

Inclusive top-quark pair production at hadron colliders has been computed 
fully differentially to next--\-to--\-next--\-to--\-leading order (NNLO) 
in the strong coupling expansion~\cite{Czakon:2013goa,Czakon:2015owf}. 
Predictions for top-quark pair production in association with up to two jets are
available at the next--\-to--leading order 
(NLO)~\cite{Dittmaier:2007wz,Bredenstein:2009aj,Bredenstein:2010rs,Bevilacqua:2009zn,Bevilacqua:2010ve,
Bevilacqua:2011aa}, and 
NLO calculations for inclusive top-quark pair production and in association with up to one or two jets were matched
to parton showers in order to provide predictions at the particle level~\cite{Frixione:2002ik,Frixione:2007nw,Kardos:2011qa,
Alioli:2011as,Frederix:2012ps,Kardos:2013vxa,Cascioli:2013era,Hoeche:2013mua,Hoeche:2014qda,Czakon:2015cla}.

In this letter we report on the first computation of top-quark pair production with up to 
three jets at NLO QCD. 
At present only few scattering processes with more than six
external legs are known at NLO~\cite{Berger:2010zx,Ita:2011wn,Bern:2013gka,Badger:2013yda,Badger:2013ava,Denner:2015yca,Bevilacqua:2015qha,Denner:2016jyo},
and the calculation at hand is the first one that deals with a 
$2\to 5$ process with seven colored external particles including also heavy quarks.
Detailed predictions are presented for $pp\to \ttbar+0,1,2,3$\,jets 
at 13\,TeV, both at the level of cross sections and differential distributions.
We also investigate the scaling behavior of $\ttbar+$multijet cross sections
with varying jet multiplicity.

The characteristic scales of $\ttbar+$multijet production, i.e.~the invariant mass of the 
$\ttbar$ system and the transverse momentum threshold
for jet production, are typically separated by more than one order of
magnitude, while differential observables involve multiple scales,
which can be distributed over more than two orders of
magnitude.  In this situation, finding renormalization and
factorization scales that ensure a decent convergence of perturbative QCD
for the widest possible range of observables is not trivial.
Moreover, in the presence of a wide spectrum of scales, the usage of
standard factor-two variations for the estimation of theoretical
uncertainties due to missing higher-order effects becomes questionable.
Motivated by these observations, to gain more insights into the scale dependence of $\ttbar+$multijet
production and related uncertainties we compare a fixed-order
calculation, with the standard scale choice $\HT /2$, against results based
on the \MINLO method~\cite{Hamilton:2012np}.
The scale $\HT/2$ was found to yield stable and reliable
NLO predictions for $V+$multijet production~\cite{Berger:2009ep},
while the \MINLO method is especially well suited for 
multi-scale QCD processes, as it controls, through next-to-leading logarithmic (NLL) resummation,
the various higher-order logarithms
that emerge from soft and collinear effects 
in the presence of widely separated scales.
The present study provides a first systematic comparison of 
the two approaches.

\section{Details of the calculation}

Our calculations are performed using the event generator 
\Sherpa~\cite{Gleisberg:2003xi,Gleisberg:2008ta} in combination with \OpenLoops~\cite{Cascioli:2011va,OLhepforge}, a 
fully automated one-loop generator based on a numerical recursion that allows 
the fast evaluation of scattering amplitudes with many external particles. 
For the reduction to scalar integrals and for the
numerical evaluation of the latter we used \CutTools~\cite{Ossola:2007ax} in
combination with \OneLOop~\cite{vanHameren:2010cp} and, alternatively, the
\Collier library~\cite{Denner:2016kdg}, which implements the methods
of~\cite{Denner:2002ii,Denner:2005nn,Denner:2010tr}.
Tree  amplitudes are computed using \Comix~\cite{Gleisberg:2008fv},
a matrix-element generator based on the color-dressed Berends-Giele recursive
relations~\cite{Duhr:2006iq}. Infrared singularities are canceled using the  
dipole subtraction method~\cite{Catani:1996vz,Catani:2002hc}, as automated in \Comix,
with the exception of K- and P-operators that are taken from the implementation 
described in~\cite{Gleisberg:2007md}. \Comix is also used for the evaluation
of all phase-space integrals. Analyses are performed with the help of
\Rivet~\cite{Buckley:2010ar}.

\begin{table}
\begin{center}
\begin{tabular}{l|rrrr}
partonic channel \textbackslash~$N$ & 0 & 1 & 2 & 3\\
\hline
$gg\to\ttbar+N\,g$ & 47 & 630 & 9'438 & 152'070 \\
$u\bar u \to\ttbar+N\,g$ & 12 & 122 & 1'608 & 23'835\\
$u\bar u \to\ttbar u\bar u+(N-2)\,g$ & -- & -- & 506 & 6'642\\
$u\bar u \to\ttbar d\bar d+(N-2)\,g$ & -- & -- & 252 & 3'321
\end{tabular}
\end{center}
\label{tab:diagcounting}
\caption{Number of one-loop Feynman diagrams in representative partonic channels in 
$pp\to \ttbar+N$\,jets for $N=0,1,2,3$.}
\end{table}

We carry out a series of $pp\to \ttbar+N$\,jet
NLO calculations with $N=0,1,2,3$, taking into account the exact dependence on the 
number of colors, $N_c=3$. As an illustration of the rapid growth of complexity 
at high jet multiplicity, in Table~\ref{tab:diagcounting} we list the 
number of one-loop Feynman diagrams that contribute 
to a few representative partonic channels. 
In addition to the presence of more than $10^5$ loop diagrams 
in the $gg\to \ttbar+3g$ channel, we note that also 
the very large number of channels not listed in 
Table~\ref{tab:diagcounting} as well as the 
computation of real contributions
pose very serious challenges in the 
\mbox{ $\ttbar+3$\,jet} calculation.

Proton--proton cross sections are obtained by using,
both at LO and NLO, the CT14 NLO PDF set~\cite{Dulat:2015mca}
with five active flavors,  and the  corresponding strong coupling. 
Matrix elements are computed with massless $b$-quarks,
and top-quarks are kept stable. Hence, 
our results can be compared to data only upon 
reconstruction of the $\ttbar$ system and extrapolation 
of fiducial measurements to the full phase space.
However, we expect the main features shown in our analysis to be
present also in computations including top-quark decays
and acceptance cuts. 
The latter
will undoubtedly play a role, 
but the reduction of scale uncertainties is generic as long as
the radiative phase space is not heavily restricted by experimental cuts. Apart from
performing a direct analysis, we also provide Root NTuples~\cite{Bern:2013zja} that can be 
used in the future for more detailed studies including top-quark decays and matching 
to parton showers. 

In our standard perturbative calculations we employ renormalization
and factorization scales  
defined as $\muR=\muF=\HT/2$, where
$\HT=\sum_i \sqrt{p_{\rT,i}^2+m_i^2}$, with the 
sum running over all (anti)top quarks and light partons, including also real radiation at NLO.
Results generated in this manner are compared to alternative computations based 
on the \MINLO procedure~\cite{Hamilton:2012np}.
To this end, we have realized 
a fully automated implementation of the 
\MINLO method in \Sherpa.

\section{\MINLO method and implementation}

The \MINLO method can be regarded as a generalized scale setting approach 
that guarantees a decent perturbative 
convergence 
for differential  multi-jet cross sections. This is achieved via
appropriate scale choices~\cite{Amati:1980ch} and Sudakov form factors~\cite{Catani:1991hj}
that resum NLL enhancements in the soft and collinear regions of phase space.
To this end, in the case of $\ttbar+$multijet production,
LO partonic events of type $ab\to \ttbar+N$\,partons 
are recursively clustered back to a core process $\abtilde\to \ttbar$
by means of a $k_\rT$ jet algorithm~\cite{Catani:1993hr}.
The resulting clustering history is interpreted as an event topology,
where the $N$-jet final state emerges from the core process
through a sequence of successive branchings that
take place at the scales 
$q_N,\ldots,q_2,q_1$
and are connected by propagators.
The nodal scales $q_i$ correspond to the $k_\rT$ measure of the jet algorithm, 
and only $1\to 2$ branchings consistent with the QCD interaction vertices are
allowed.
In our implementation of the 
$k_\rT$ jet algorithm we use the definition of $\Delta R$ 
given in Eq.~(11) of~\cite{Catani:1993hr}
and we set $\Delta R=0.4$.
Typically, the $k_\rT$ algorithm gives rise 
to ordered branching histories with
$q_1<\dots<q_N<\mucore$, where $\mucore$ is the characteristic hard scale of the
core process. However, also unordered branchings can occur. 
For instance, this can happen in the presence of jets with transverse momenta above 
$\mucore$.
Since soft-collinear resummation does not make sense for such hard emissions,
in our \MINLO implementation 
possible unordered clusterings are undone and alternative
ordered configurations are considered.
At the end, the branching history is restricted to  
ordered branchings $q_1<\dots<q_{\tN}<\mucore$, where
$\tN=N-M$. The remaining $M$ jets that can not be clustered 
in an ordered way are treated as part of the core process,
and $\mucore$ is evaluated according to the kinematics of the
corresponding $\ttbar+M$\,jet hard event.

At LO, the renormalization scale $\muR$ is chosen according to
the event branching history in such a way that 
\begin{equation}\label{eq:muR}
\left[\as(\mu_{R})\right]^{N+2} = \left[\as(\mucore)\right]^{2+M} \prod_{i=1}^{\tN} \as(q_i),
\end{equation}
and in our calculation we set $\mucore=\HT/2$.

The resummation of soft and collinear logarithms is achieved by dressing external and internal 
lines of the event topology by Sudakov form factors.
At variance with the original formulation of \MINLO~\cite{Hamilton:2012np},
in our implementation we employ the symmetry of the LO DGLAP splitting 
functions, $P_{ab}(z)$, to define physical Sudakov form factors
\begin{equation}\label{eq:def_nll_sudakov}
  \begin{split}
    &\Delta_a(Q_0,Q)=\exp\left\{-\int_{Q_0}^Q\frac{{\rm d} q}{q}
    \frac{\as(q)}{\pi}\sum_{b=q,g}\right.\\
    &\quad\left.\int_0^{1-q/Q}{\rm d}z
    \left(z\,P_{ab}(z)+\delta_{ab}\frac{\as(q)}{2\pi}\frac{2C_a}{1-z}K\right)\right\}\;,
  \end{split}
\end{equation}
where~\cite{Catani:1990rr}
\begin{equation}
  K=\left(\frac{67}{18}-\frac{\pi^2}{6}\right)C_A-\frac{10}{9}T_R\,n_f\;,
\end{equation}
and $a=g,q$ corresponds to massless gluons and quarks, respectively.
The representation~\eqref{eq:def_nll_sudakov}
allows the interpretation of $\Delta_a(Q_0,Q)$ in terms of no-branching probabilities
between the scales $Q_0$  and $Q$.

Given a LO event topology
with $\tN$ ordered branchings,
the lowest branching scale,
$\qmin=q_1$, is identified as resolution scale,
and  
the $\tN$ emissions are supplemented by Sudakov form factors
that render them exclusive w.r.t.~any extra emissions above $\qmin$.
This is achieved by dressing each external line of flavor $a=q,g$
connected with the $i$-th  branching 
by a form factor $\Delta_{a}(\qmin,q_i)$, while internal lines that 
connect successive branchings $k<l$ are dressed by factors
$\Delta_{a}(\qmin,q_l)/$ $\Delta_{a}(\qmin,q_k)$, which correspond to 
no-branching probabilities between $q_k$ and $q_l$ at resolution scale $\qmin$.
For internal lines that connect branchings at $q_k$ to the core process 
analogous no-branching probabilities
between $q_k$ and $\mucore$ are applied.
Sudakov form factors along the incoming lines
provide a NLL resummation 
that corresponds to the evolution of PDFs
from the resolution scale $\qmin$ 
to the hard scale of the core process. 
Therefore, for consistency, PDFs are evaluated at the factorization scale $\muF=\qmin$.

The generalization to NLO requires only two straightforward modifications of the LO
algorithm. First, for what concerns the scale setting and 
Sudakov form factors, the contributions that live in the $N$-parton phase space, i.e.~Born and one-loop
contributions as well as all IR-subtraction terms, are handled exactly as in LO.
Instead, real-emission events that lead to histories with 
$\tN+1\le N+1$ ordered branchings
at scales $q_0<q_1<\dots <q_{\tN}$  are handled as Born-like $\tN$-parton
events with resolution scale
$\qmin=q_1$, i.e.~the softest branching at the scale $q_0$ is considered as unresolved and is simply
excluded from the \MINLO procedure. In other words, the softest emission at NLO is not 
dressed with Sudakov form factors and does not enter the definitions of 
$\muR$ and  $\muF$.
Second, appropriate counterterms are introduced in order to subtract the
overall $\mathcal{O}(\as)$ contribution from  Sudakov form factors,
such as to avoid double counting of NLO effects.

Concerning the treatment of top quarks a few extra comments are in order. 
Given the low rate at which top quarks radiate jets, such emissions are
simply neglected in our implementation of the \MINLO procedure 
by excluding top quarks from the clustering algorithm.
To quantify the uncertainty arising from this approach,
we implemented an alternative algorithm that allows the combination of top quarks with other 
final-state partons in the massive Durham scheme~\cite{Krauss:2003cr,Rodrigo:2003ws}.
The difference between the two procedures is found to be about 10\% at leading order and
5\% at next-to-leading order for the observables studied here, and it is therefore smaller 
than the renormalization and factorization scale uncertainties.
Finally, also the top quarks that enter the core process are dressed with 
Sudakov form factors $\Delta_t(\qmin,\mucore)$, 
which render them exclusive w.r.t.~emissions above 
$\qmin$.
To compute the Sudakov form factors $\Delta_t$,
we include quark masses in the splitting functions, 
according to the method 
described in~\cite{Krauss:2003cr,Rodrigo:2003ws}, using the corresponding extension
of Eq.~\eqref{eq:def_nll_sudakov}. This means 
in particular that we use the massive splitting functions from~\cite{Catani:2000ef},
the propagator corrections listed in~\cite{Krauss:2003cr,Rodrigo:2003ws}, 
and we replace the two-loop cusp term $K\,2C_F/(1-z)$ by $K\,C_F(2/(1-z)-m^2/p_ip_j)$ 
in the case of massive quark splittings $\widetilde{\imath\jmath}\to i,j$.

Scale uncertainties in the \MINLO framework are assessed through standard 
factor-two variations of $\muR$ and $\muF$. 
The renormalization scale is kept fixed in the
Sudakov form factors but is varied as usual in the rest of the (N)LO
cross section, including the counterterms that subtract the 
$\mathcal{O}(\as)$ parts of the Sudakov form factors at NLO.
Variations $\muF\to \xiF\,\muF$ of the factorization scale 
are more subtle.
 They have to be applied at the level of PDFs and related NLO counterterms, as well 
as in the Sudakov form factors that depend on $\qmin=\muF$. More precisely,
$\qmin\to \xiF\,\qmin$ variations are applied only to Sudakov form factors associated with external and internal
initial-state lines, 
and Sudakov form factors $\Delta_a(\xiF\,\qmin,q_k)$ are set to one 
when $\xiF\,\qmin$ exceeds $q_k$.

\section{Predictions for the 13\,TeV LHC}

In the following we present selected predictions
for $pp\to \ttbar+0,1,2,3$\,jets at 13\,TeV.
We construct jets by clustering light partons with the anti-$k_t$  
algorithm~\cite{Cacciari:2008gp} at $R=0.4$, 
and by default we select jets with pseudorapidity \mbox{$|\eta_{\jet}|<2.5$}
and a jet-$\pT$ threshold of $25$\,GeV.
Unless stated otherwise, depending on the minimum number $N$ of jets that is required by the 
observable at hand,
inclusive (N)LO or MI(N)LO calculations 
with $N$ jets are used.

\begin{figure}[]
\begin{minipage}{0.49\textwidth}
\begin{center}
  \includegraphics[scale=0.62,trim=0 25 10 0,clip]{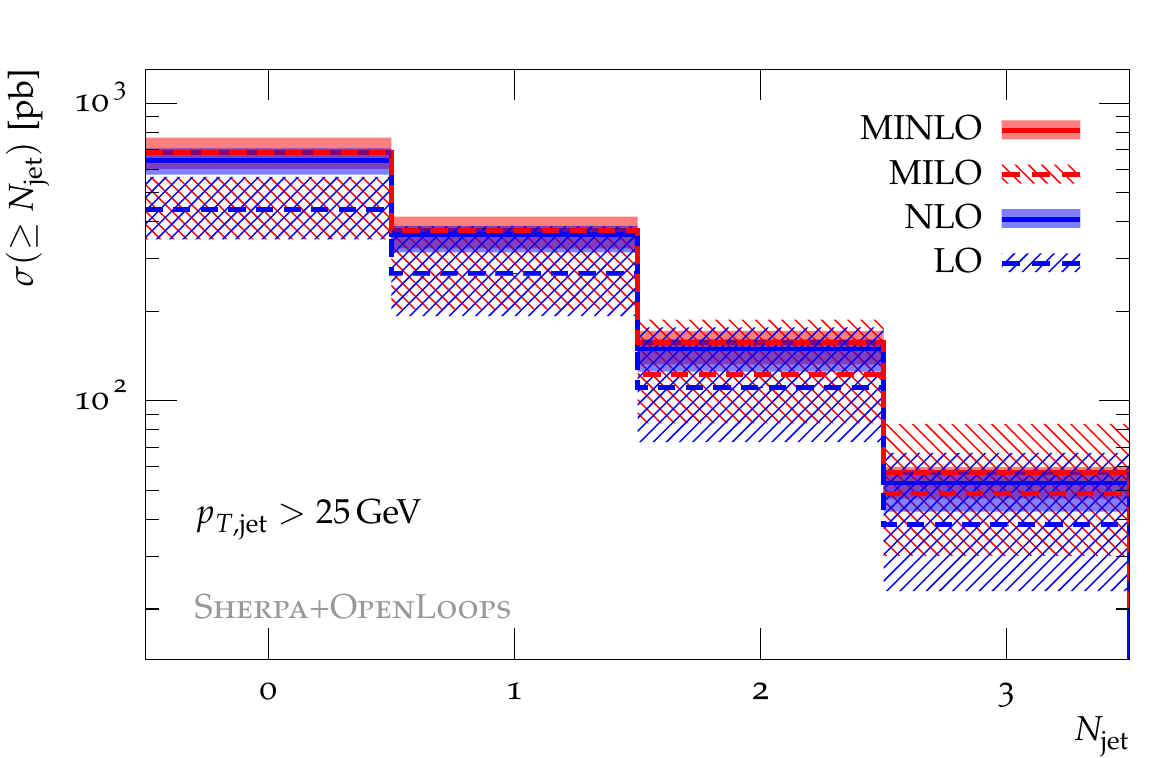}\\[-4pt]
  \includegraphics[scale=0.62,trim=0 25 10 0,clip]{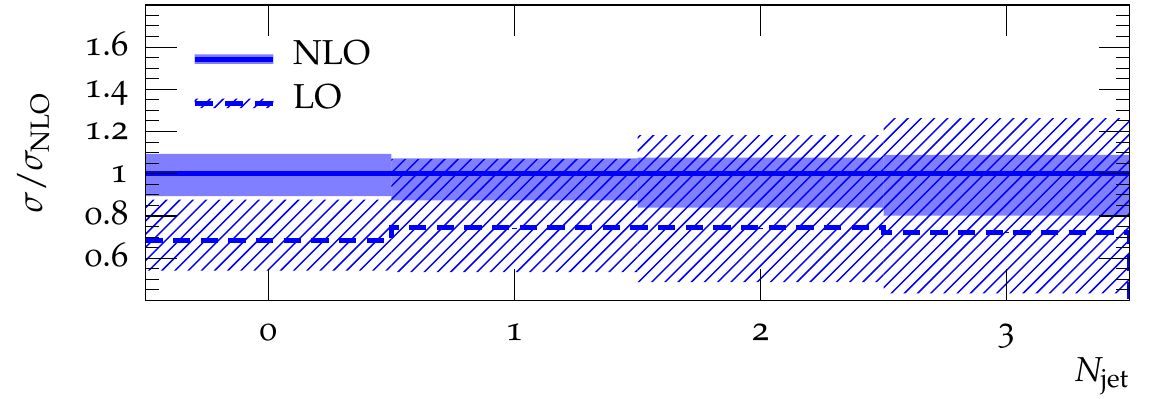}\\[-4pt]
  \includegraphics[scale=0.62,trim=0 25 10 0,clip]{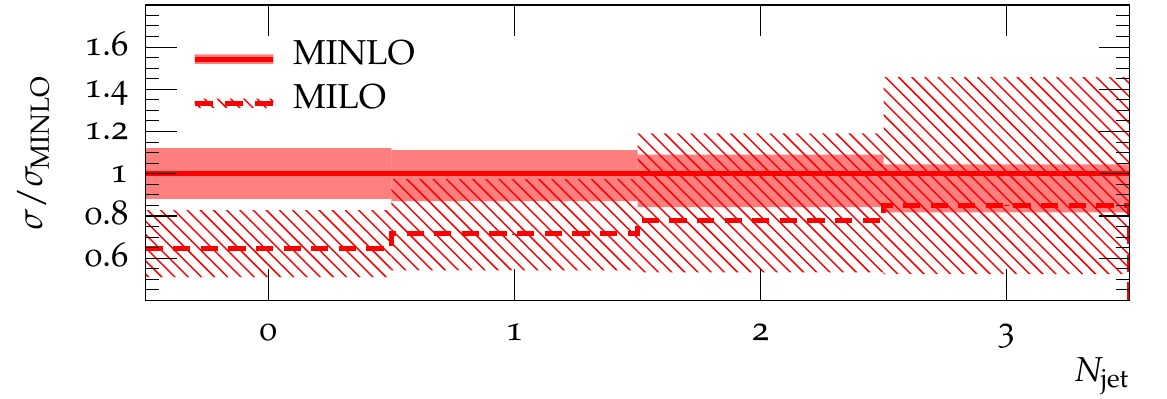}\\[-4pt]
  \includegraphics[scale=0.62,trim=0 0 10 0,clip]{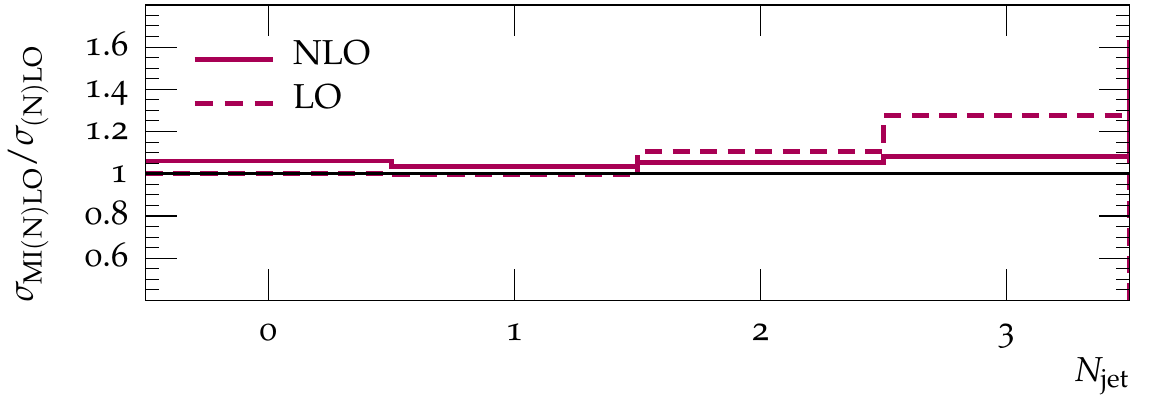}
\end{center}
\end{minipage}
\caption{\label{fig:njet_bsp}
Inclusive $\ttbar+$multijet cross sections with a minimum number 
$N=0,1,2,3$ of jets at $\ptjet\ge$25\,GeV. See the main text for details.}
\end{figure}
\begin{figure}[]
\begin{minipage}{0.49\textwidth}
\begin{center}
  \includegraphics[scale=0.62,trim=0 25 10 0,clip]{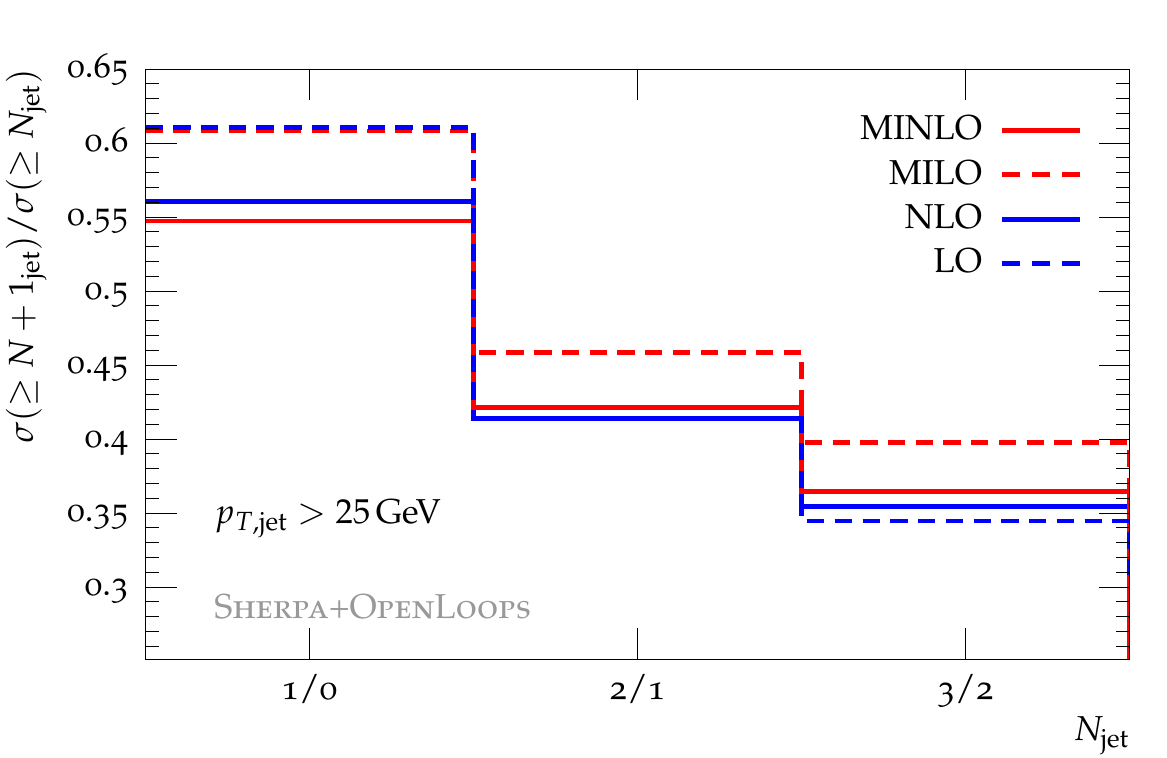}\\[-4pt]
  \includegraphics[scale=0.62,trim=0 25 10 0,clip]{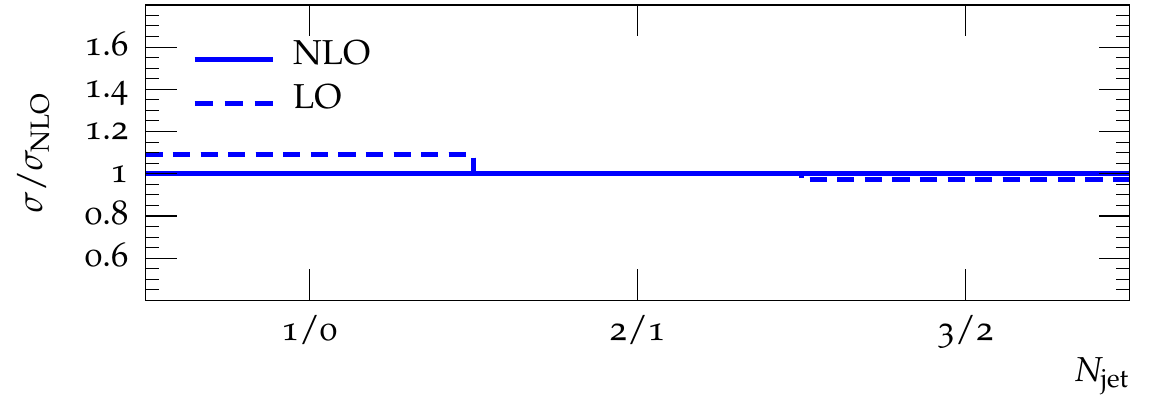}\\[-4pt]
  \includegraphics[scale=0.62,trim=0 25 10 0,clip]{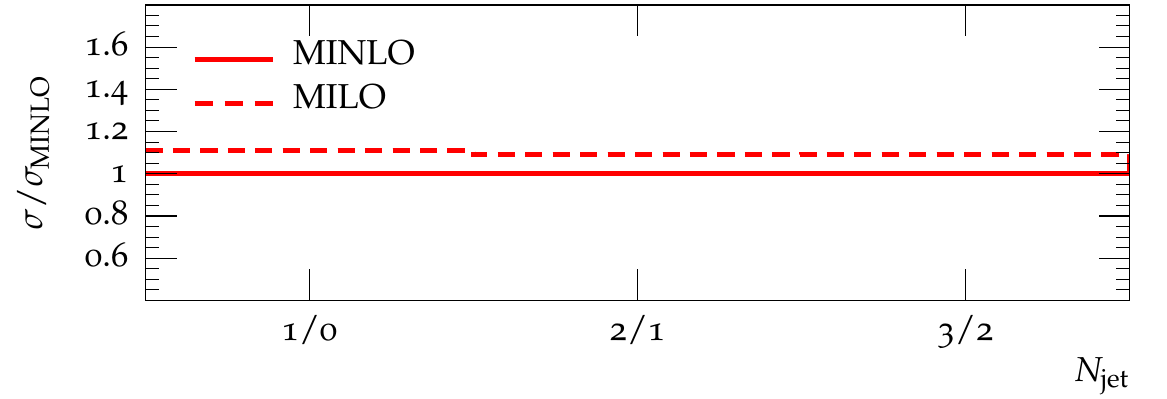}\\[-4pt]
  \includegraphics[scale=0.62,trim=0 0 10 0,clip]{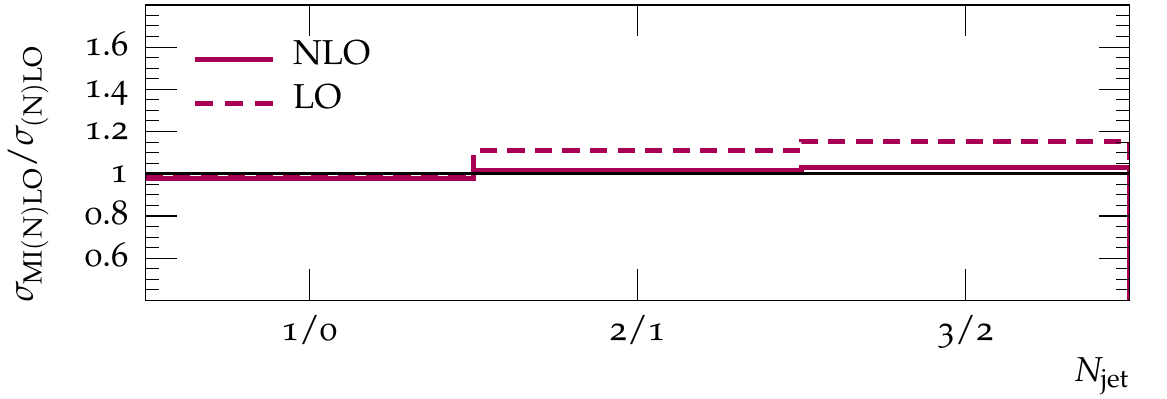}
\end{center}
\end{minipage}
\caption{\label{fig:njet_ratios}
Ratios of $\ttbar+N$\,jet
over $\ttbar+(N-1)$\,jet inclusive cross sections
for $N=1,2,3$ and $\ptjet\ge$25\,GeV.
}
\end{figure}

The jet multiplicity distribution is presented in~\reffi{fig:njet_bsp}.
The top panel displays four predictions, stemming from fixed-order LO and NLO calculations, 
and from \MINLO computations 
at LO and NLO (labeled `MILO' and `MINLO'). The second panel shows the ratio between
LO and NLO predictions at fixed order, 
while the third panel
shows the ratio between MILO and MINLO predictions.
The last panel
shows the ratio between MINLO and NLO. 
The bands illustrate scale uncertainties 
estimated through independent factor-two rescaling of 
$\muR$ and $\muF$ excluding antipodal variations.
Fixed-order predictions feature rather large NLO corrections of about $+50\%$ for all 
jet multiplicities, while \MINLO results feature steadily decreasing 
corrections for increasing $\njets$. In both cases, LO scale uncertainties tend to grow
by more than $10\%$ at each extra jet emission, while
(MI)\-NLO scale uncertainties are significantly reduced and the total width of 
the (MI)NLO variation bands is about 20--25\% for all considered $\njets$ values.
Comparing fixed-order NLO and MINLO predictions
we observe a remarkable agreement at the level of 4--8\%.
This supports NLO and MINLO scale-uncertainty estimates based on factor-two variations and
encourages the usage of either 
of the two calculations (NLO and MINLO) in practical applications.

\begin{table*}[h]
\include{table_INC_all}
\caption{\label{tab:incxsptall}
Inclusive ($\njets\ge n$) and exclusive ($\njets=n$)  cross sections with $n=0,1,2,3$ jets 
and different transverse momentum thresholds, $\ptjet\ge 25, 40, 60, 80$ GeV. 
Uncertainties represent the envelope of the independent $\muR$ and $\muF$ 
variations around the central value (antipodal variations excluded).}
\end{table*}

As demonstrated in Table~\ref{tab:incxsptall},
the good agreement between fixed-order NLO and MINLO results and the 
consistency of the observed NLO--MINLO differences with factor-two scale variations
persist also for a range of other commonly used $\ptjet$-thresholds~\cite{ATLAS-CONF-2015-065}.
More precisely, for 
inclusive $\ttbar+N$\,jet cross sections with jet-$\pT$ thresholds of 25, 40, 65 and \mbox{80\,GeV},
MINLO predictions lie between 
5\% and 19\% above NLO ones. The largest differences are observed at large jet multiplicity and for 
large $\pT$-thresholds, in which case \MINLO cross sections feature significantly better 
perturbative convergence and smaller scale uncertainties as compared to fixed-order ones.
In Table~\ref{tab:incxsptall} also exclusive cross sections with exactly $N$ jets are presented.
In that case, the difference between MINLO and
NLO predictions varies between -7\% and +11\%. Apart from the zero-jet case, where the \MINLO
approach is not well motivated, the MINLO/NLO ratio is almost independent of the number of jets
and grows from 0.95 to 1.10 when the $\pT$-threshold increases from 25 to 80\,GeV.
Similarly as in the inclusive case, at $\pT$-thresholds above 40\,GeV
 MINLO predictions  for exclusive $N$-jet cross sections with $N\ge 2$
feature much better convergence and 
smaller scale uncertainties w.r.t.~fixed order. 
However, for lower $\pT$-thresholds the opposite is observed, and
in the three-jet case the MINLO scale uncertainty becomes twice as large at the NLO one.
This can be attributed to the fact that Sudakov logarithms related to the vetoing of 
NLO radiation are not resummed in the \MINLO approach.
In spite of this caveat, the general agreement of fixed-order NLO and 
\MINLO results remains remarkably good for all considered observables.

Figure~\ref{fig:njet_ratios} shows ratios of 
inclusive $\ttbar+N$\,jet cross sections for successive jet multiplicities.
Due to the cancellation of various sources of experimental and theoretical uncertainties,
such ratios are ideally suited for precision tests of QCD.
Corresponding ratios have been widely studied in 
vector-boson plus multi-jet production~\cite{Bern:2014fea,Bern:2014voa},
where a striking scaling behavior was observed at 
high jet multiplicity.
In the case of \mbox{$\ttbar$+multijet} ratios involving up to three jets
we find a moderate dependence on the number of jets but
no clear scaling.
This behavior is rather similar to scaling violations in $V+$\,multijet
production at lower multiplicity and, analogously as for $V+$\,multijets, 
can be attributed to the suppression of important partonic channels in the zero-jet process at LO.
In fact, quark--gluon channels are not active in $t\bar{t}$ production at LO.
In addition,  at LHC energies the gluonic initial state is strongly favored due to the
parton luminosity and the $t$-channel enhancement of the $gg\to\ttbar$ cross section, such that
the situation becomes similar to vector boson production, except for the difference of 
quark versus gluon initial states at LO. 
When adding additional jets, firstly quark--gluon 
initial states and secondly quark--quark initial states (including $t$-channel top-quark diagrams)
are added, which contribute sizably to the cross section at larger invariant mass and/or
transverse momentum. 
In order to test scaling hypotheses, it would therefore ultimately 
be necessary to compute the $t\bar{t}+4$ jet over $t\bar{t}+3$ jet ratio, and eventually
the $t\bar{t}+5$ jet over $t\bar{t}+4$ jet ratio. This is out of reach of present technology, 
therefore we do not investigate the scaling behavior in more detail.
Nevertheless, given the excellent agreement between  
MINLO and NLO predictions up to three jets, the ratios in \reffi{fig:njet_ratios} 
can be regarded as optimal benchmarks for precision tests.

\begin{figure*}[h] 
  \begin{center}\hskip 2mm
    \includegraphics[scale=0.3999,trim=0 25 10 0,clip]{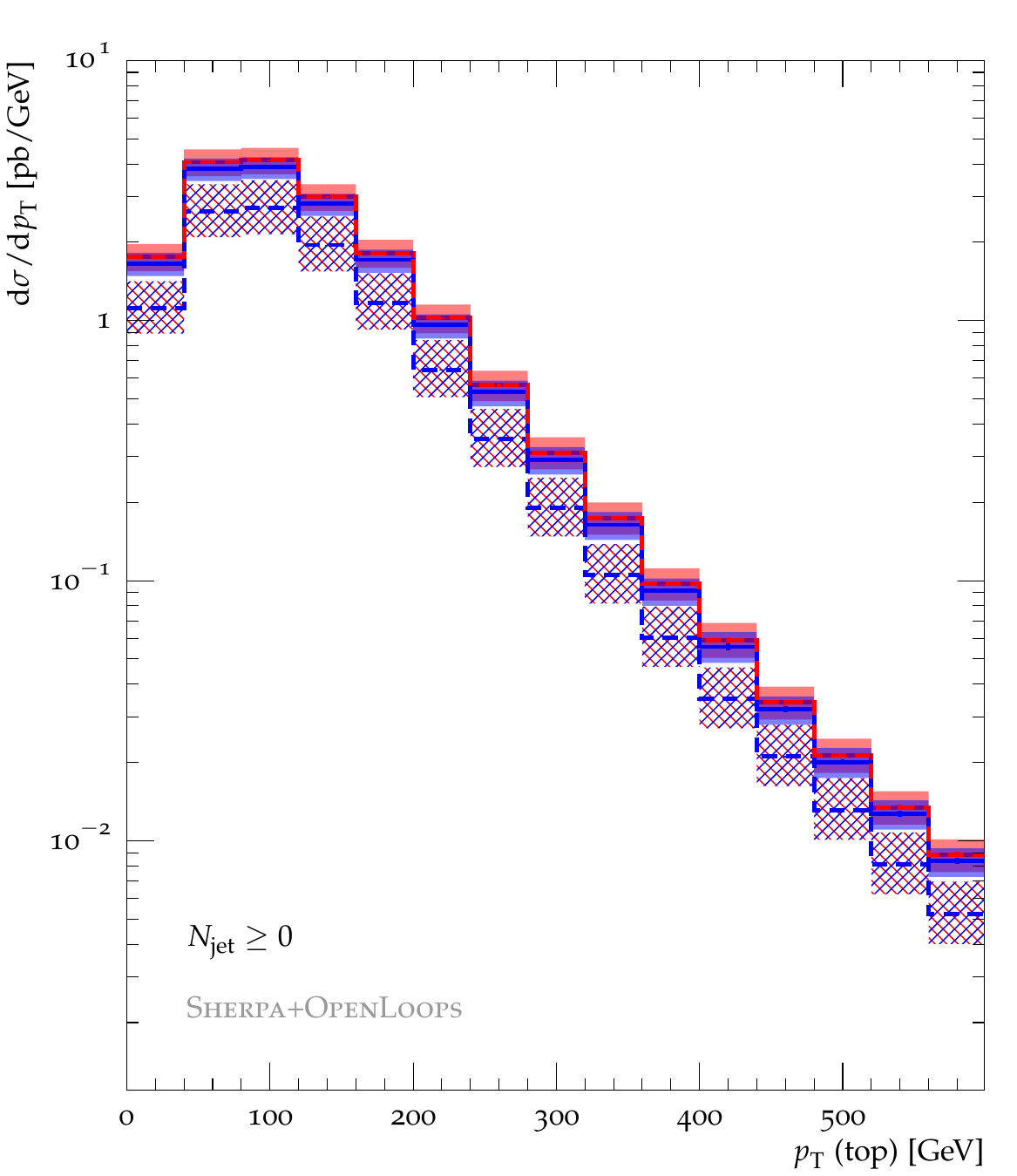}\hskipfour
    \includegraphics[scale=0.3999,trim=40 25 10 0,clip]{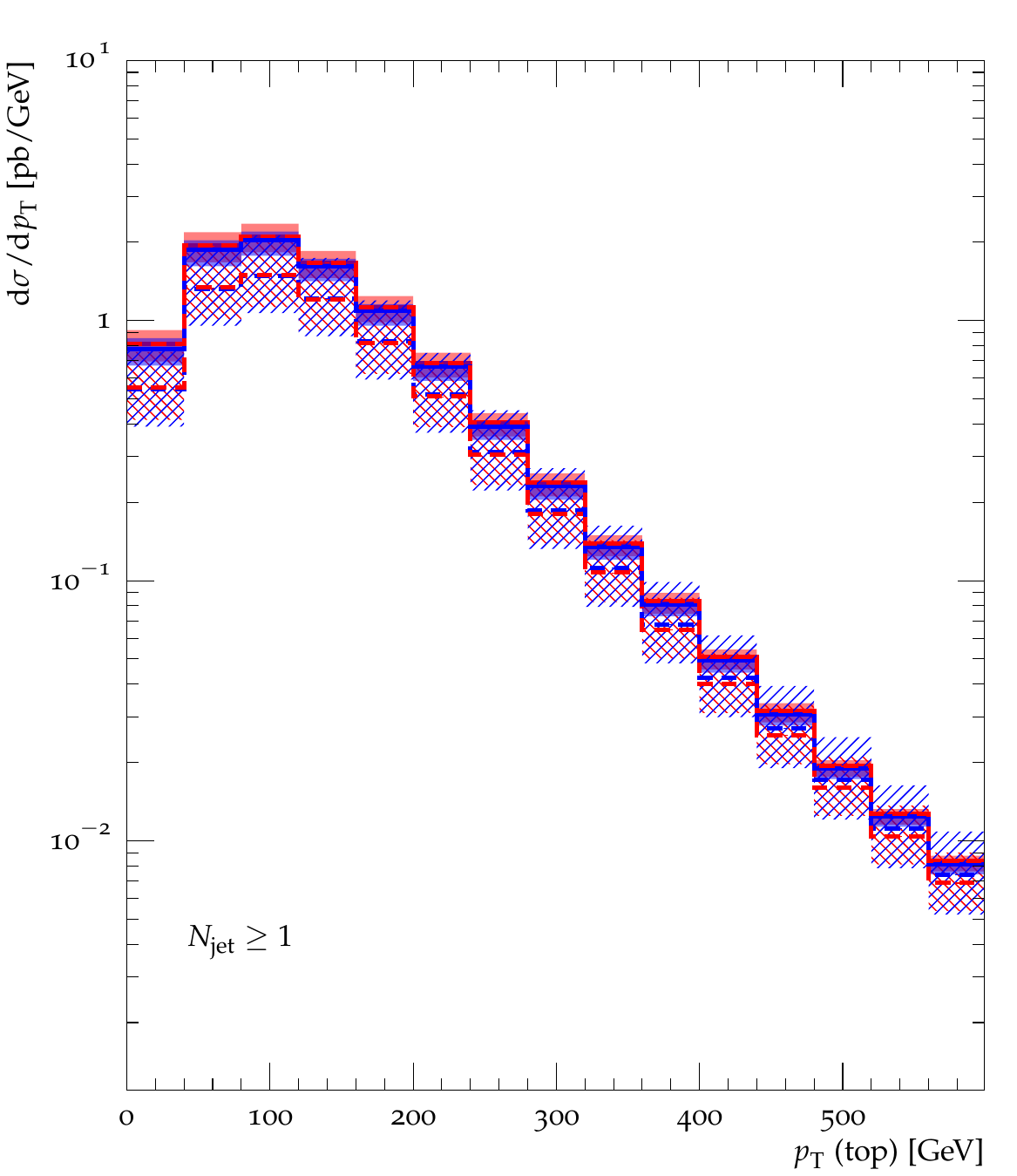}\hskipfour
    \includegraphics[scale=0.3999,trim=40 25 10 0,clip]{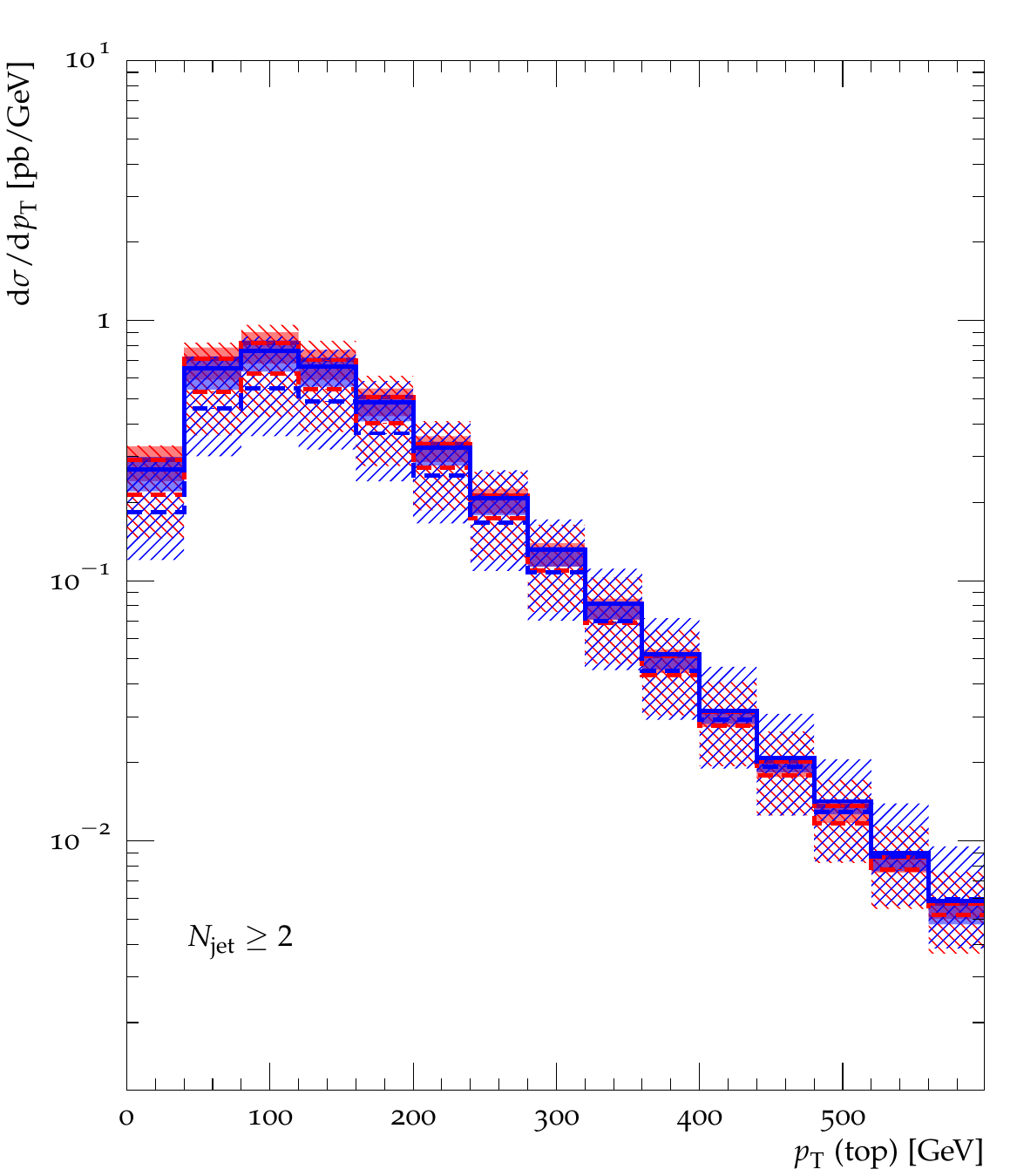}\hskipfour
    \includegraphics[scale=0.3999,trim=40 25 10 0,clip]{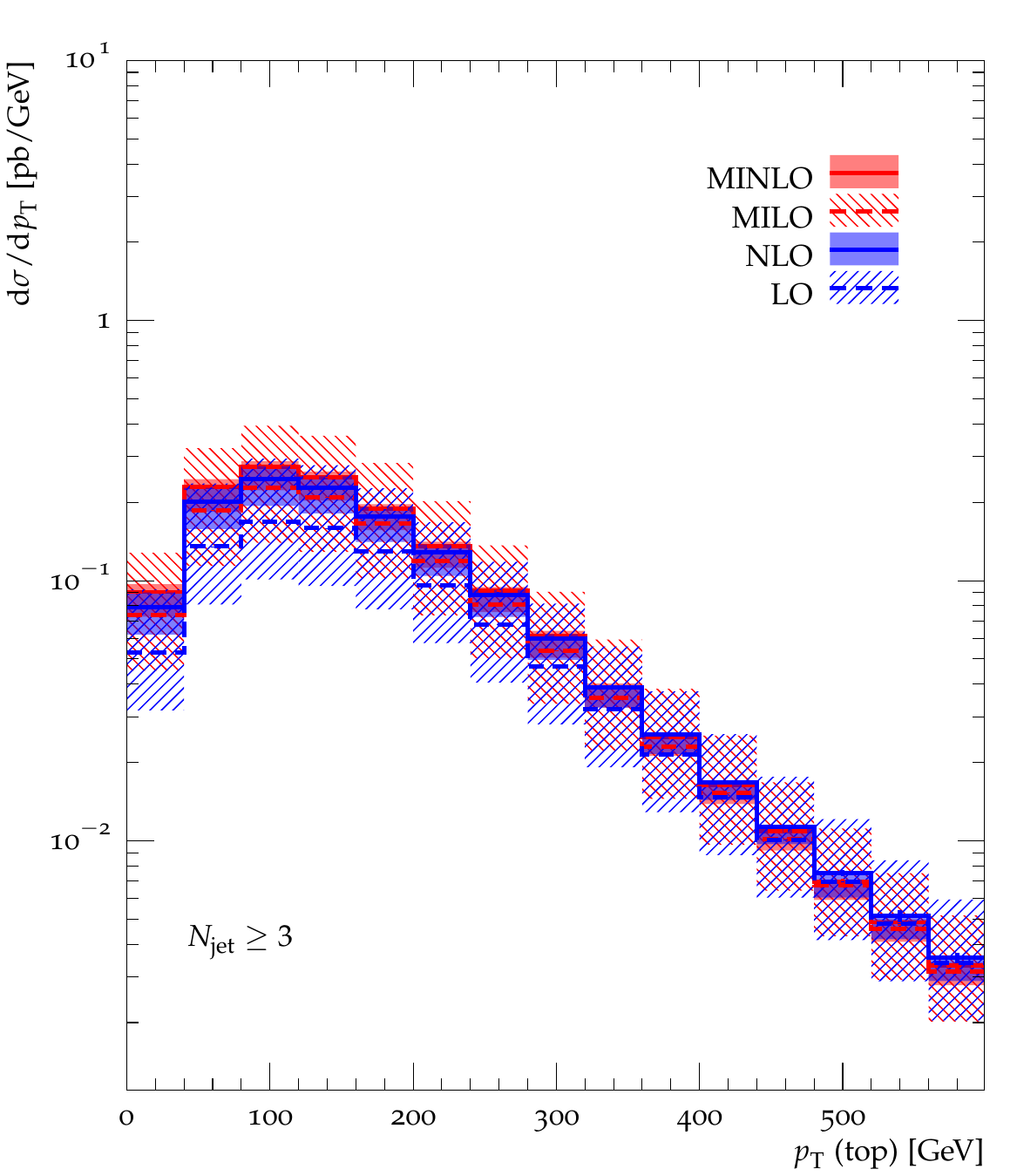}\\[-3pt]\hskip 2mm
    \includegraphics[scale=0.3999,trim=0 25 10 0,clip]{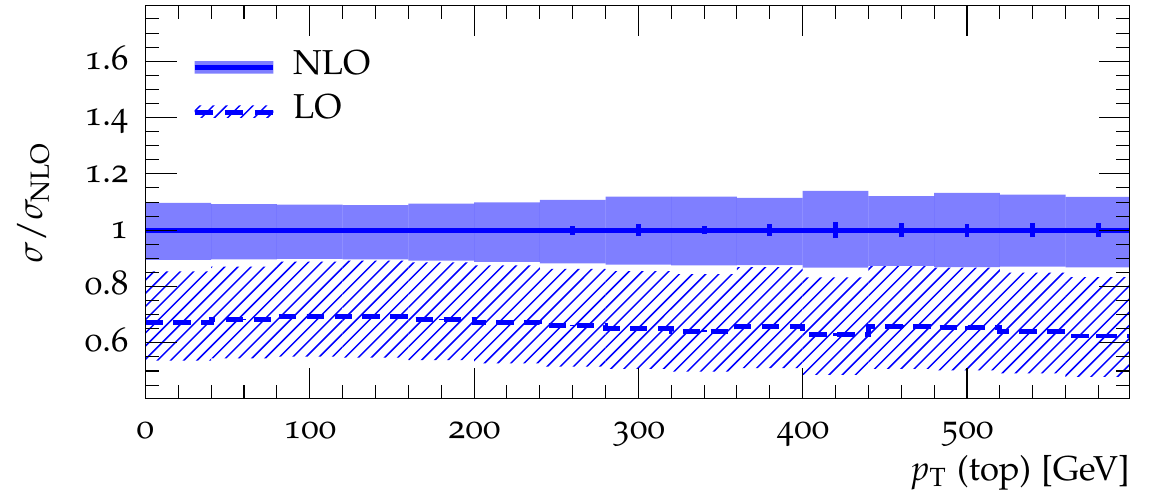}\hskipfour
    \includegraphics[scale=0.3999,trim=40 25 10 0,clip]{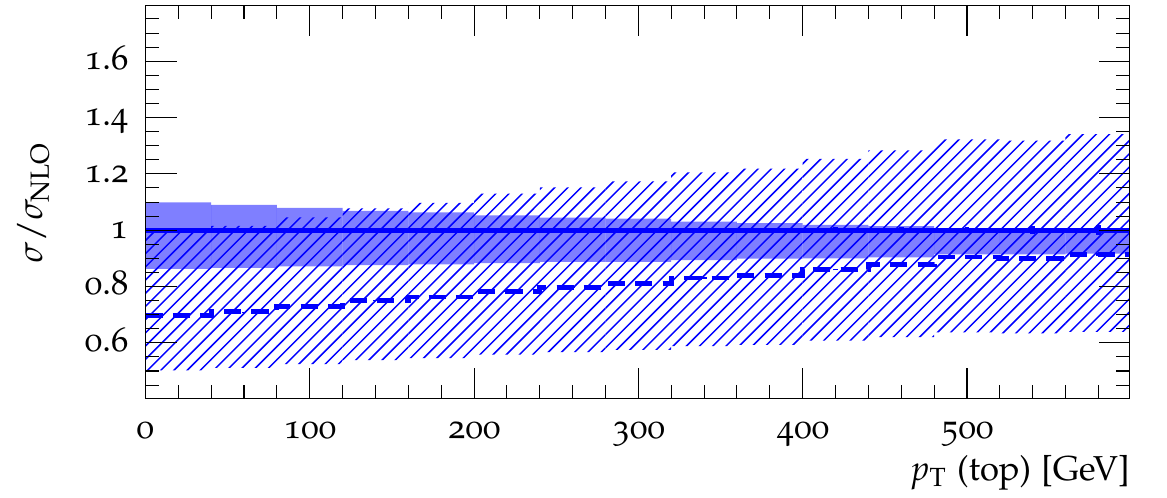}\hskipfour
    \includegraphics[scale=0.3999,trim=40 25 10 0,clip]{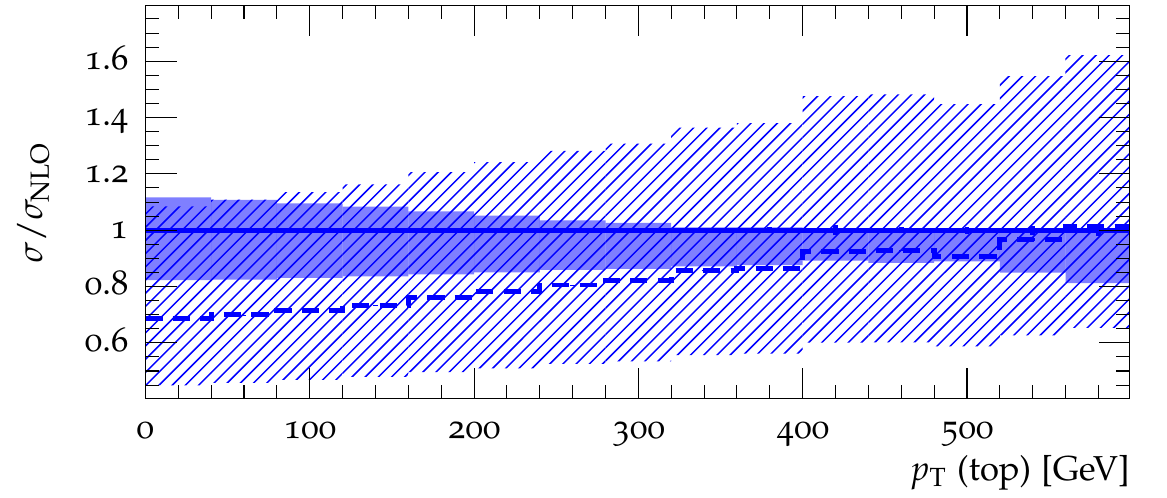}\hskipfour
    \includegraphics[scale=0.3999,trim=40 25 10 0,clip]{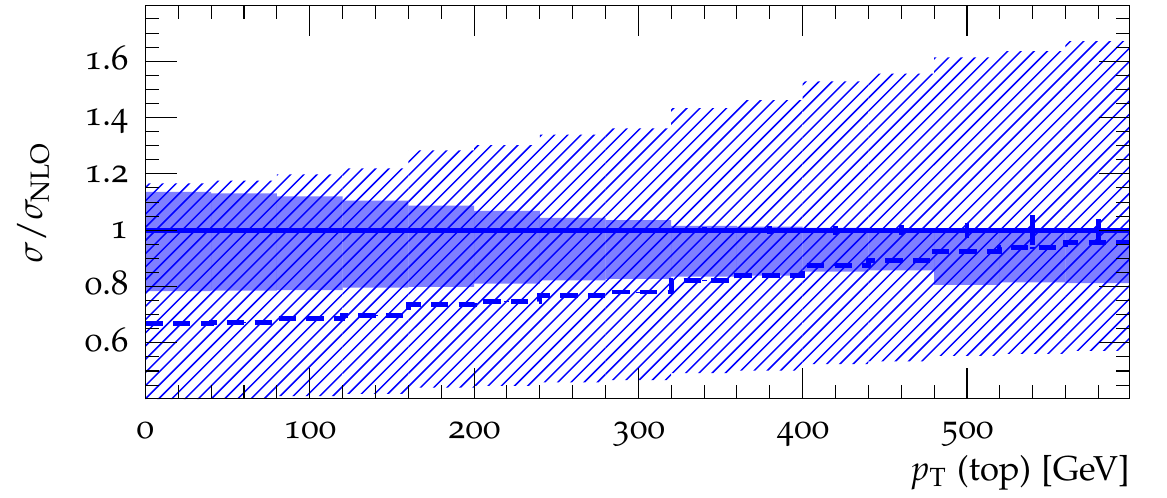}\\[-3pt]\hskip 2mm
    \includegraphics[scale=0.3999,trim=0 25 10 0,clip]{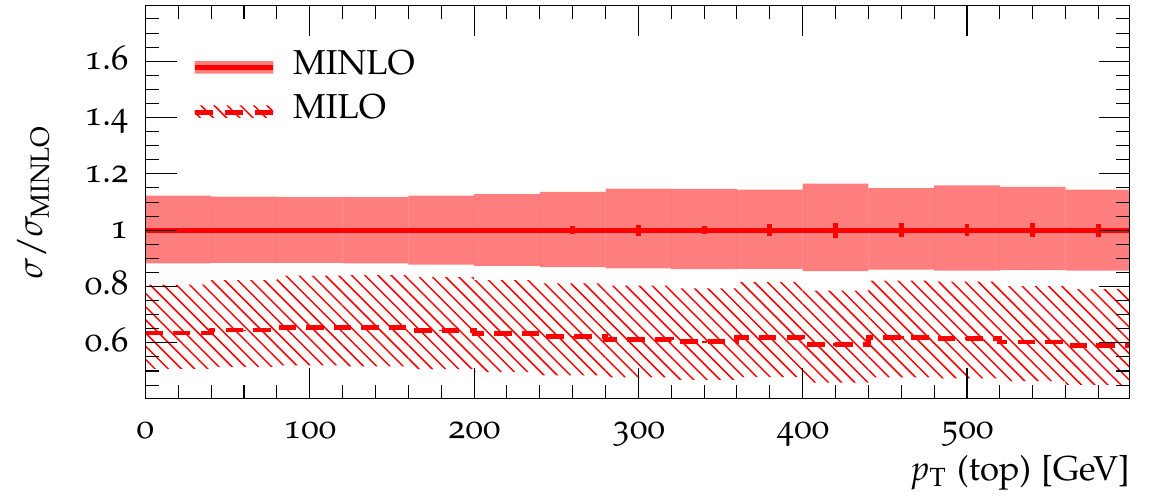}\hskipfour
    \includegraphics[scale=0.3999,trim=40 25 10 0,clip]{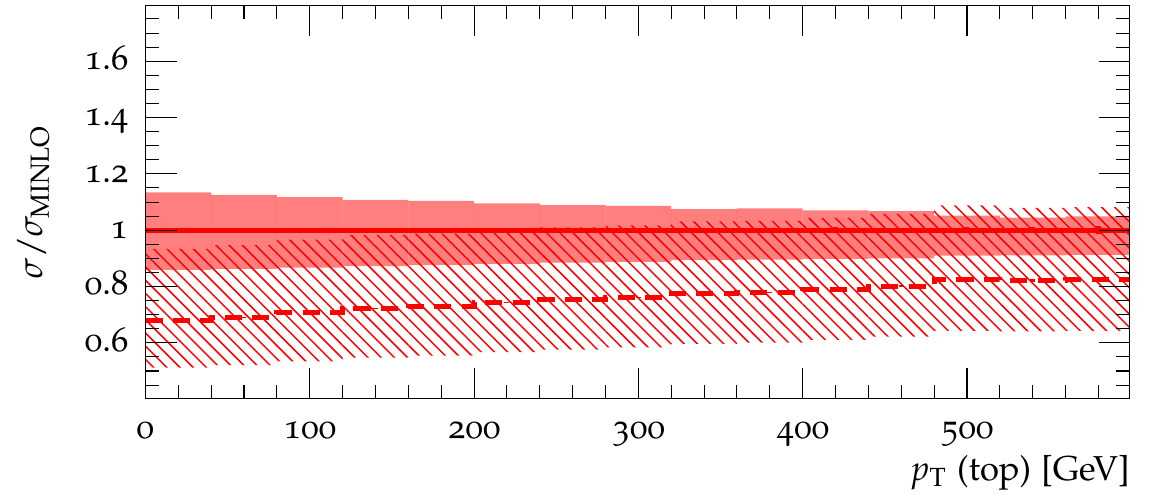}\hskipfour
    \includegraphics[scale=0.3999,trim=40 25 10 0,clip]{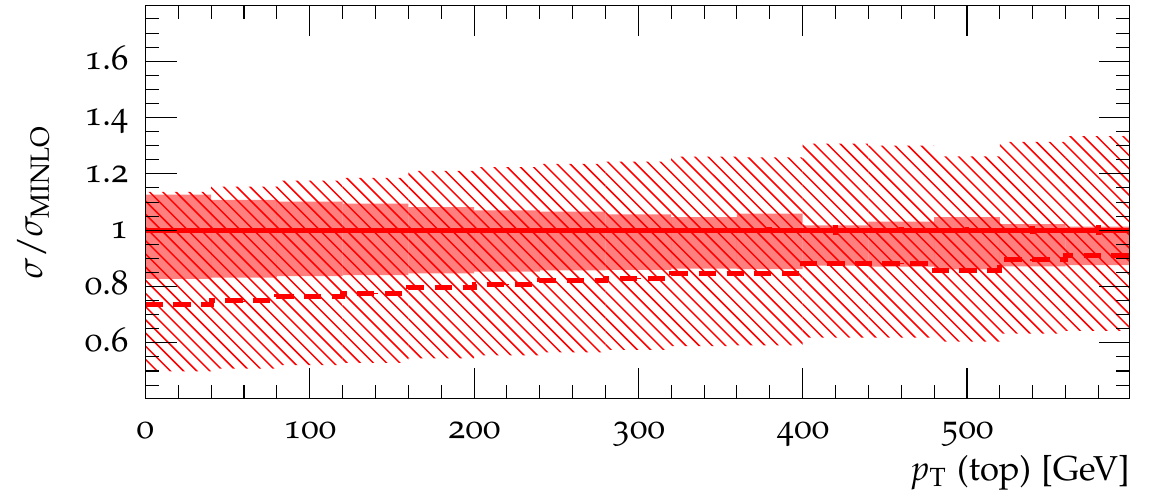}\hskipfour
    \includegraphics[scale=0.3999,trim=40 25 10 0,clip]{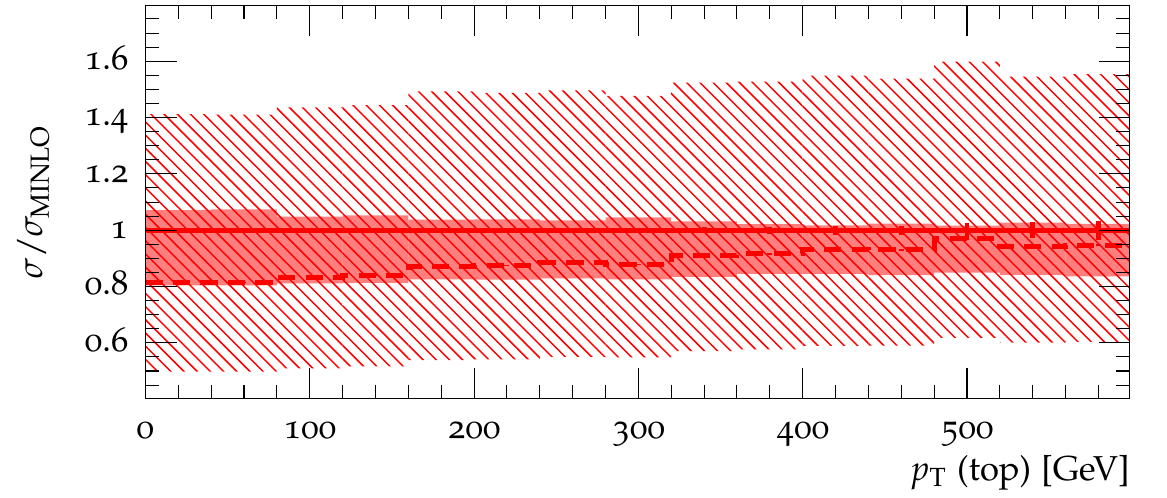}\\[-3pt]\hskip 2mm
    \includegraphics[scale=0.3999,trim=0 0 10 0,clip]{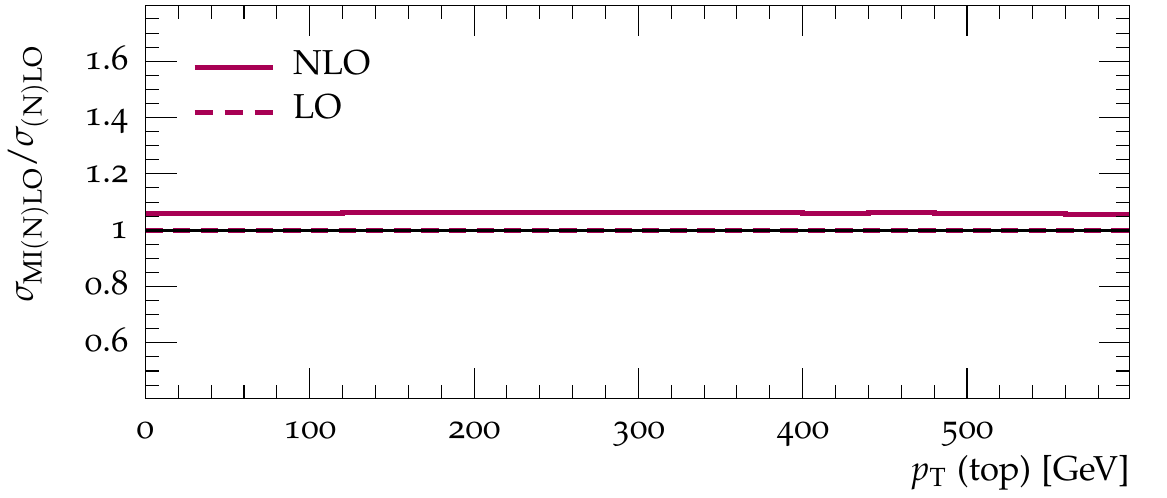}\hskipfour
    \includegraphics[scale=0.3999,trim=40 0 10 0,clip]{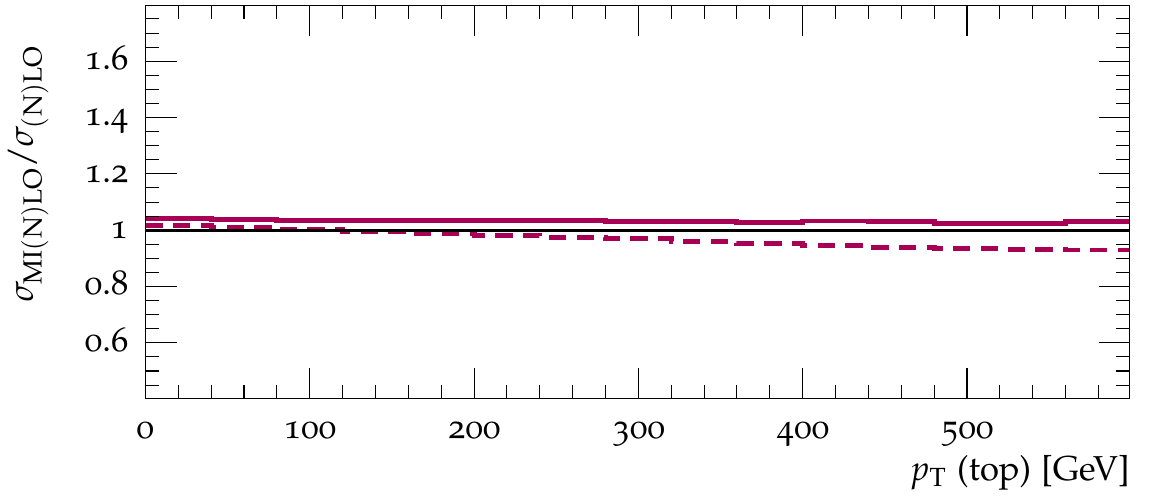}\hskipfour
    \includegraphics[scale=0.3999,trim=40 0 10 0,clip]{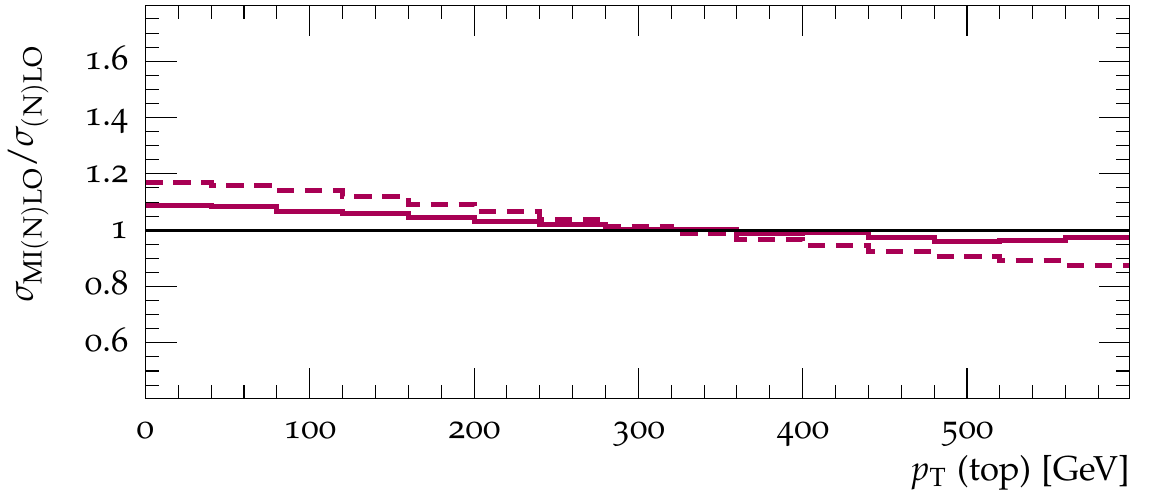}\hskipfour
    \includegraphics[scale=0.3999,trim=40 0 10 0,clip]{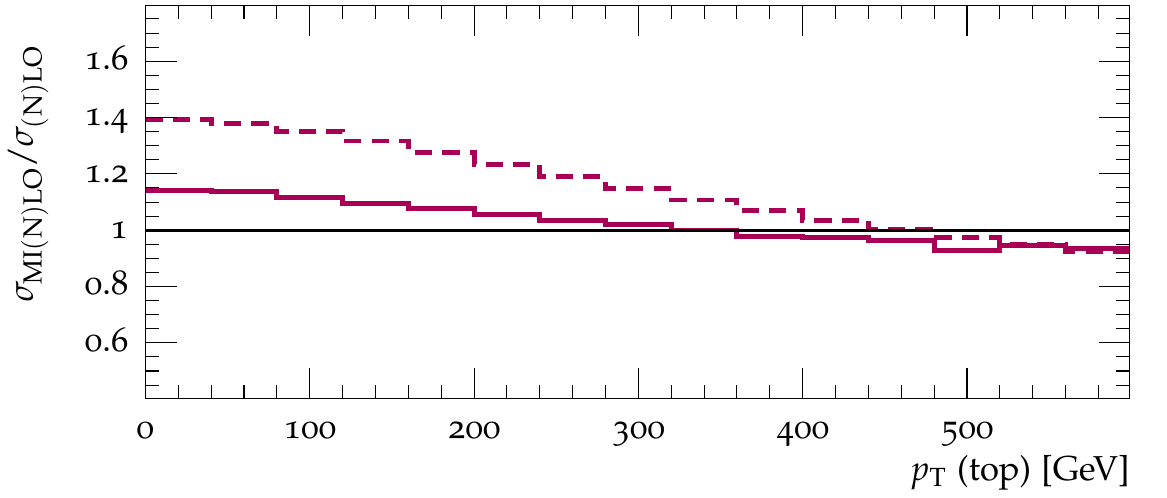}\\[3mm]
  \end{center}\capskip
  \caption{\label{fig:top_pt}
    Distribution in the top-quark $\pT$ for $pp\to \ttbar+0,1,2,3$\,jets with  $\ptjet\ge 25$\,GeV.}
\end{figure*}
\begin{figure*}[h] 
  \begin{center}
    \includegraphics[scale=0.525,trim=0 25 10 0,clip]{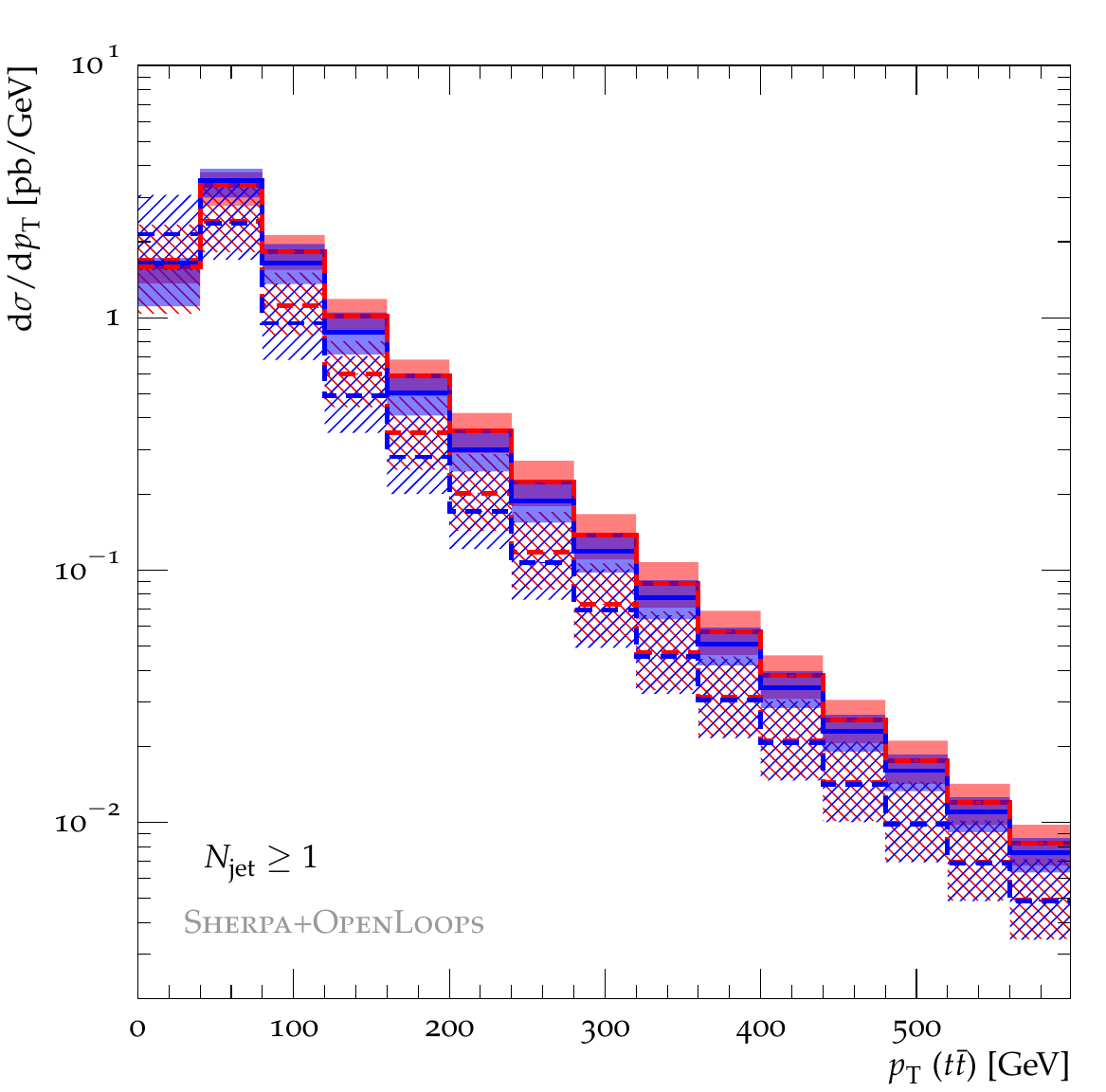}\hskipthree
    \includegraphics[scale=0.525,trim=40 25 10 0,clip]{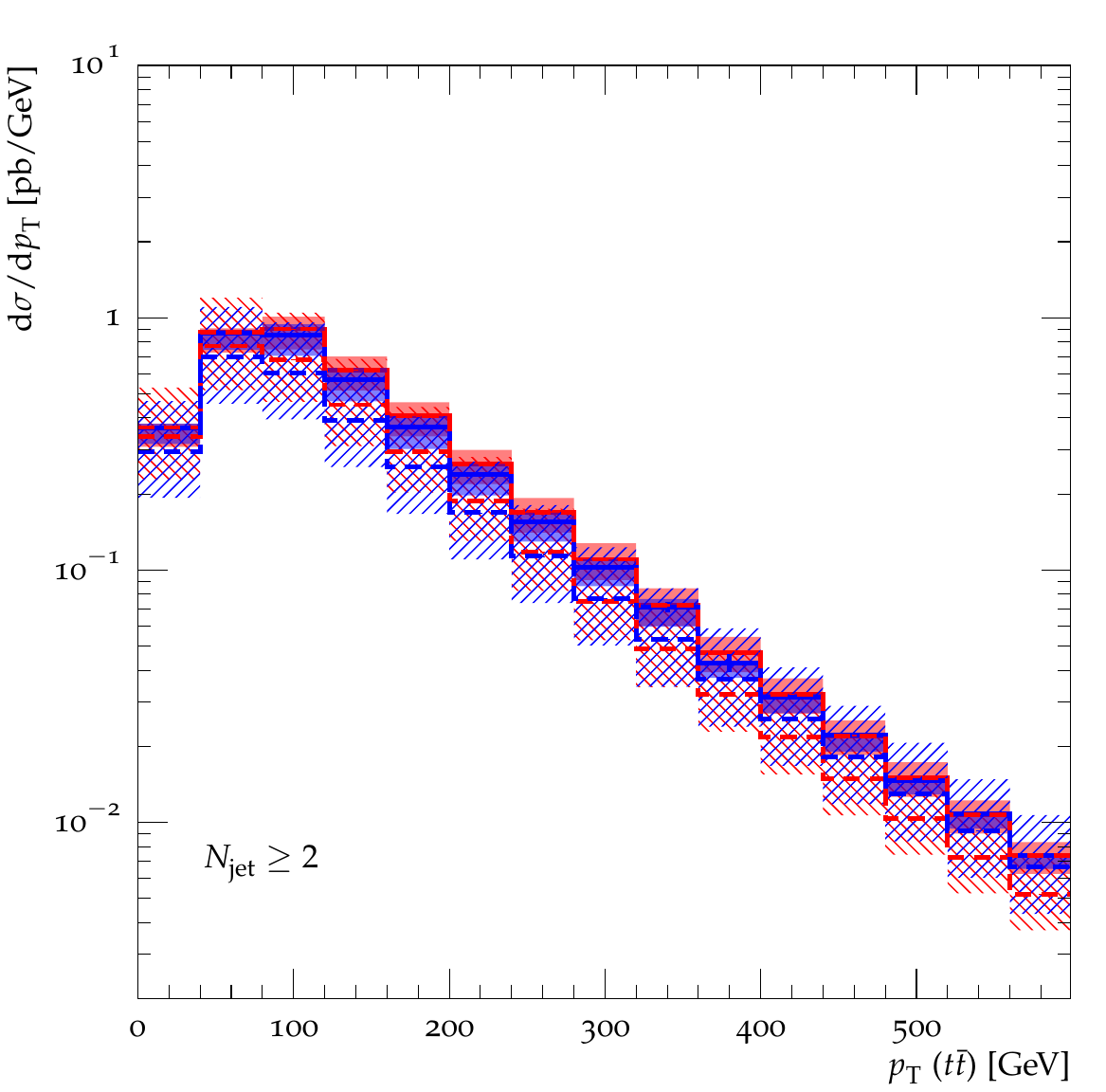}\hskipthree
    \includegraphics[scale=0.525,trim=40 25 10 0,clip]{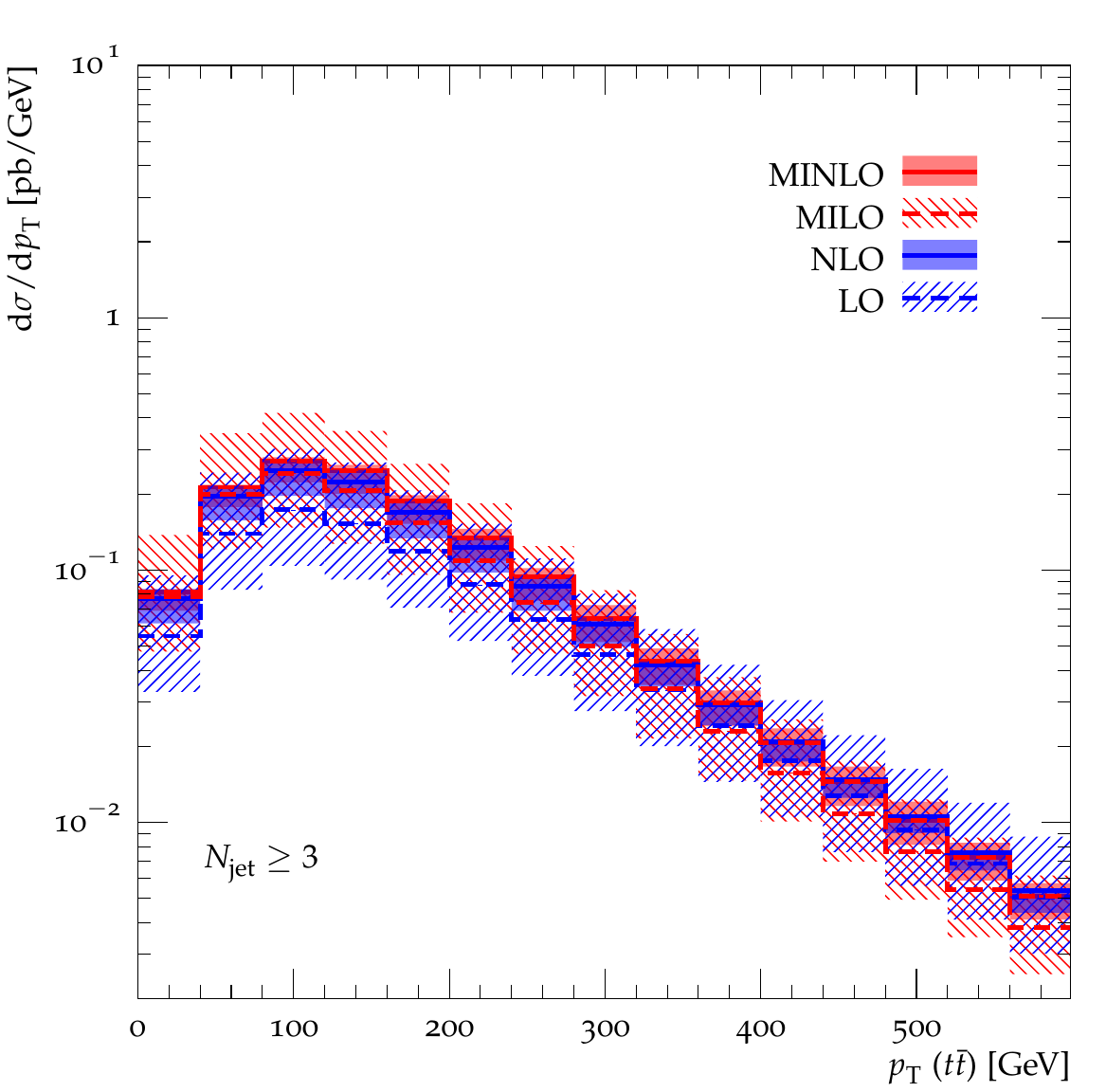}\\[-3.675pt]
    \includegraphics[scale=0.525,trim=0 25 10 0,clip]{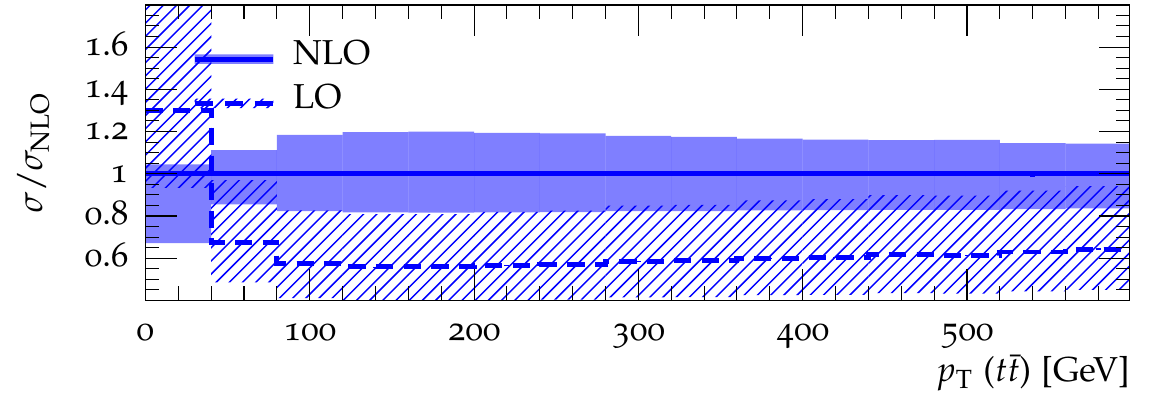}\hskipthree
    \includegraphics[scale=0.525,trim=40 25 10 0,clip]{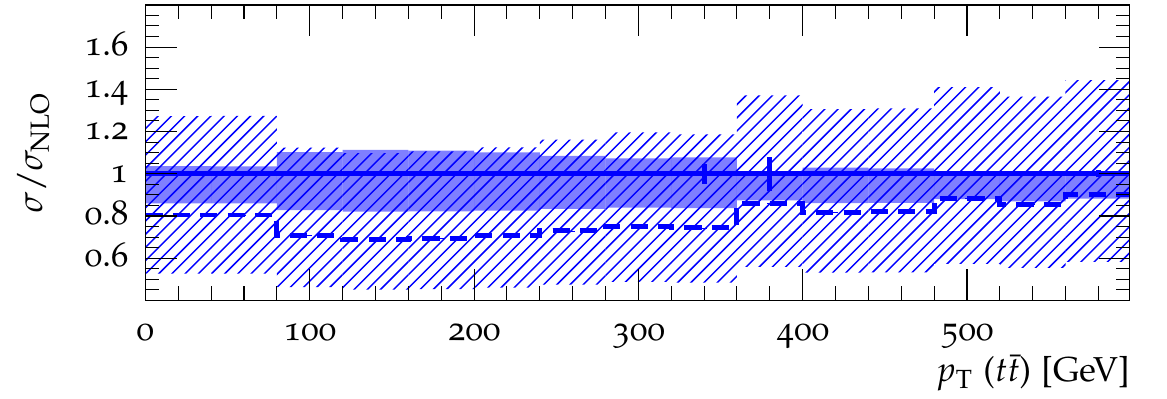}\hskipthree
    \includegraphics[scale=0.525,trim=40 25 10 0,clip]{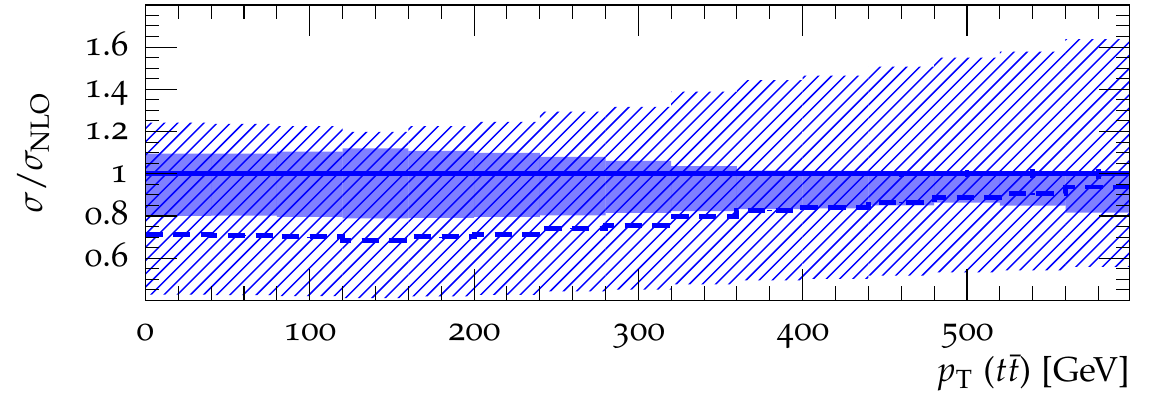}\\[-3.675pt]
    \includegraphics[scale=0.525,trim=0 25 10 0,clip]{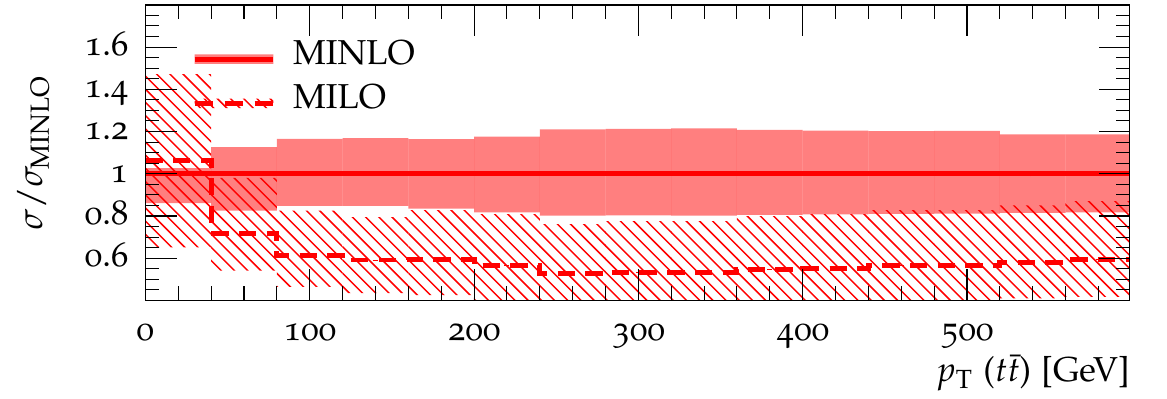}\hskipthree
    \includegraphics[scale=0.525,trim=40 25 10 0,clip]{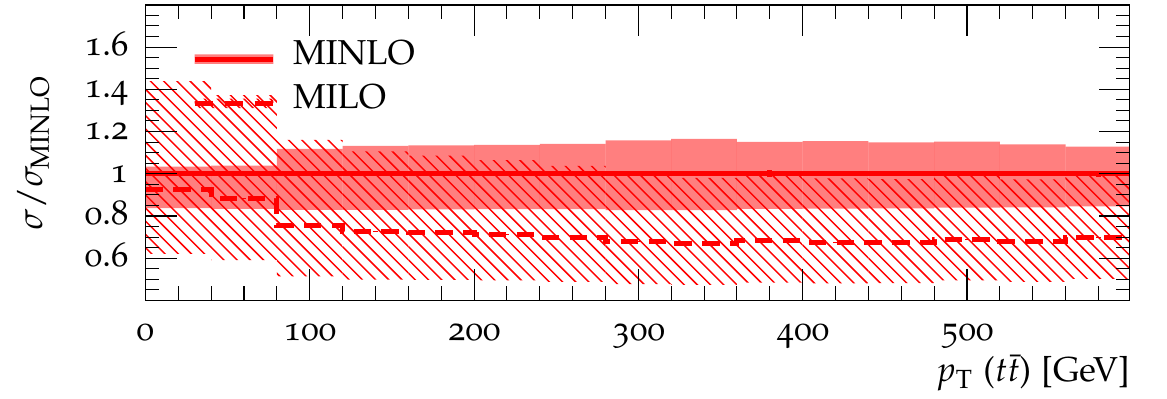}\hskipthree
    \includegraphics[scale=0.525,trim=40 25 10 0,clip]{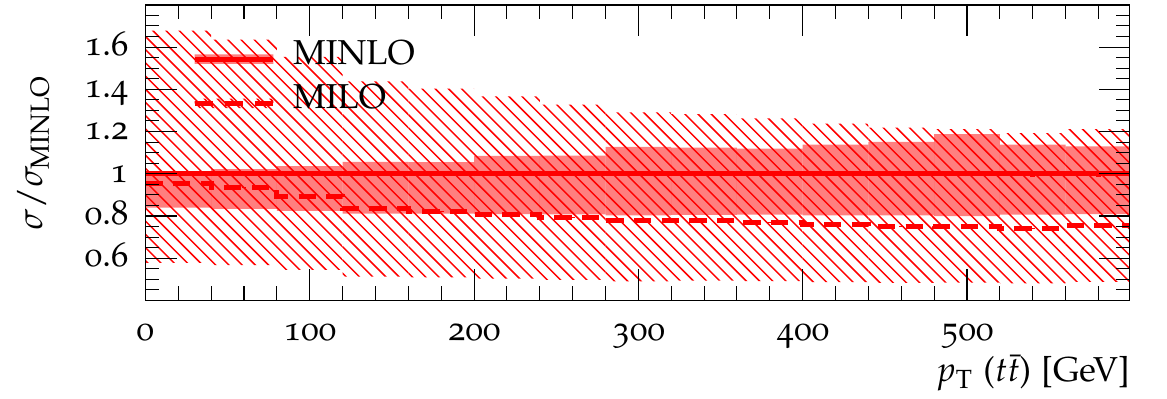}\\[-3.675pt]
    \includegraphics[scale=0.525,trim=0 0 10 0,clip]{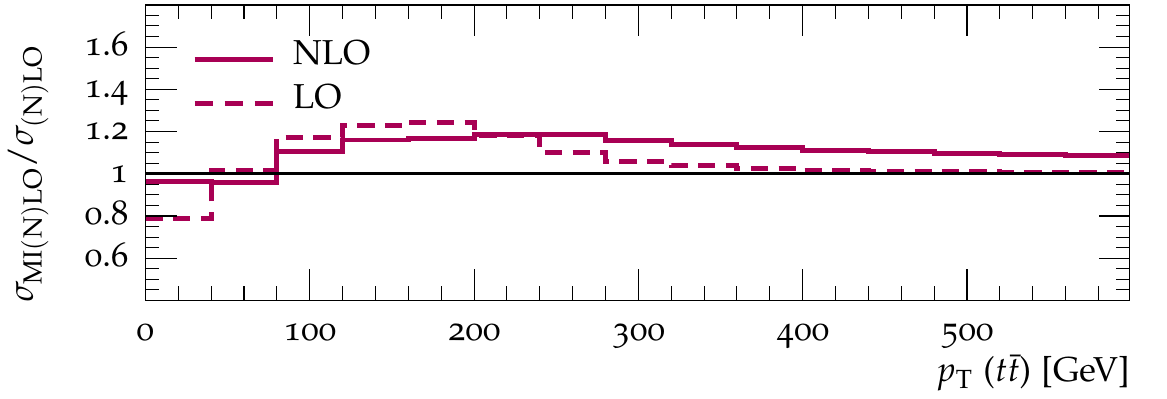}\hskipthree
    \includegraphics[scale=0.525,trim=40 0 10 0,clip]{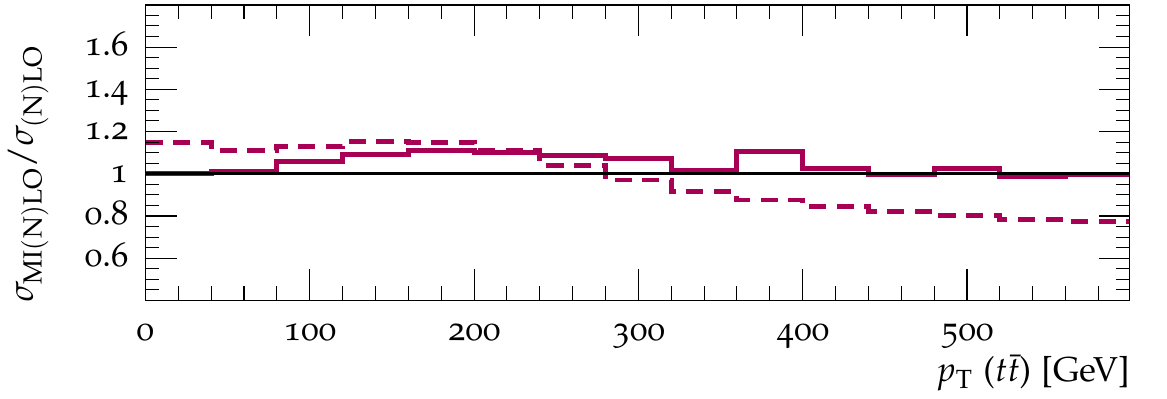}\hskipthree
    \includegraphics[scale=0.525,trim=40 0 10 0,clip]{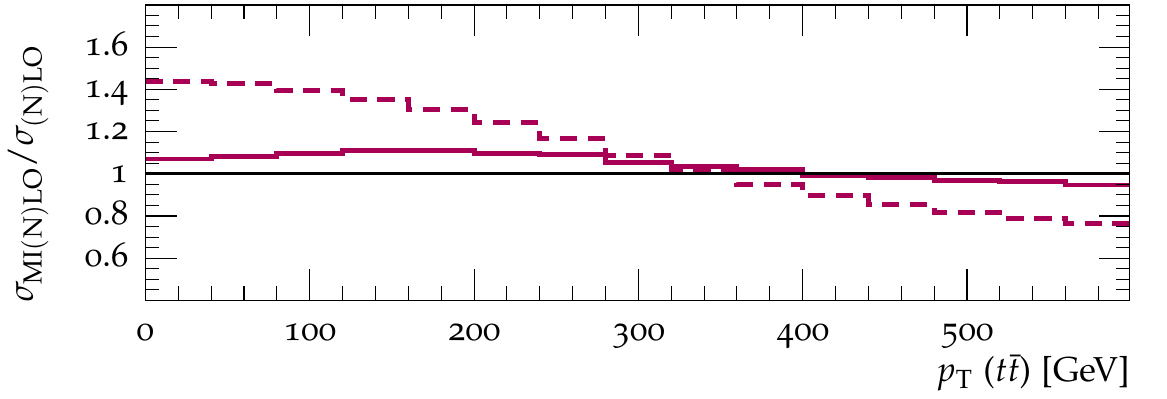}
  \end{center}\capskip
  \caption{\label{fig:ttbar_pt}
    Distribution in the $\pT$ of the $\ttbar$ system for $pp\to \ttbar+1,2,3$\,jets with $\ptjet\ge 25$\,GeV.}
\end{figure*}

Figure~\ref{fig:top_pt} shows the transverse momentum spectrum of the top quark for varying
jet multiplicities. From low to very high $\pT$ NLO scale uncertainties remain at a similarly 
small level as for integrated cross sections. For $\njets\ge 1$, we observe  
significant shape corrections, which tend to decrease at high jet multiplicity in \MINLO,
while in fixed order they remain important.
We also observe a shape difference between
fixed-order and \MINLO predictions, which tends to increase with increasing jet multiplicity
but is clearly reduced at NLO.
The overall agreement 
between fixed-order NLO and MINLO 
results is quite good, both in shape and normalization, with differences that lie within the individual scale 
uncertainties.
Figure~\ref{fig:ttbar_pt} shows the top-quark pair transverse momentum spectrum in 1-, 2- 
and 3-jet samples.
We observe a large increase in the cross section between LO and NLO in the one-jet case,
where the effect of additional radiation not modeled by the LO calculation is largest.
At higher jet multiplicities correction effects tend to decrease.
Fixed-order NLO uncertainties are similarly small as in~\reffi{fig:top_pt}, while MINLO scale uncertainties 
tend to be more pronounced in the tails.
However, we find a very good overall agreement between fixed-order NLO 
and MINLO predictions, especially for $\njets\ge 2$ and 3.

\begin{figure*}[p]
  \begin{center}
    \includegraphics[scale=0.525,trim=0 25 10 0,clip]{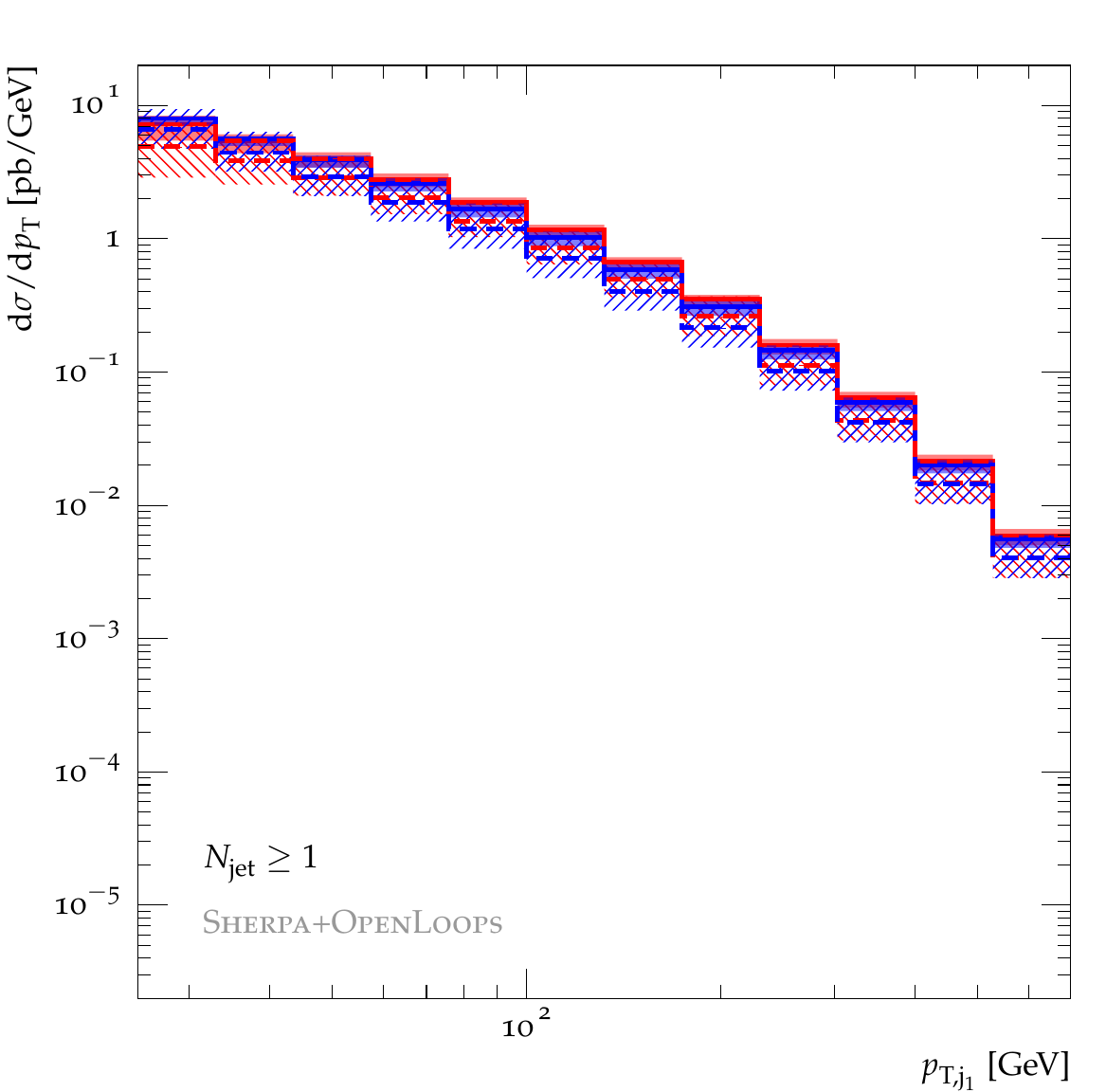}\hskipthree
    \includegraphics[scale=0.525,trim=40 25 10 0,clip]{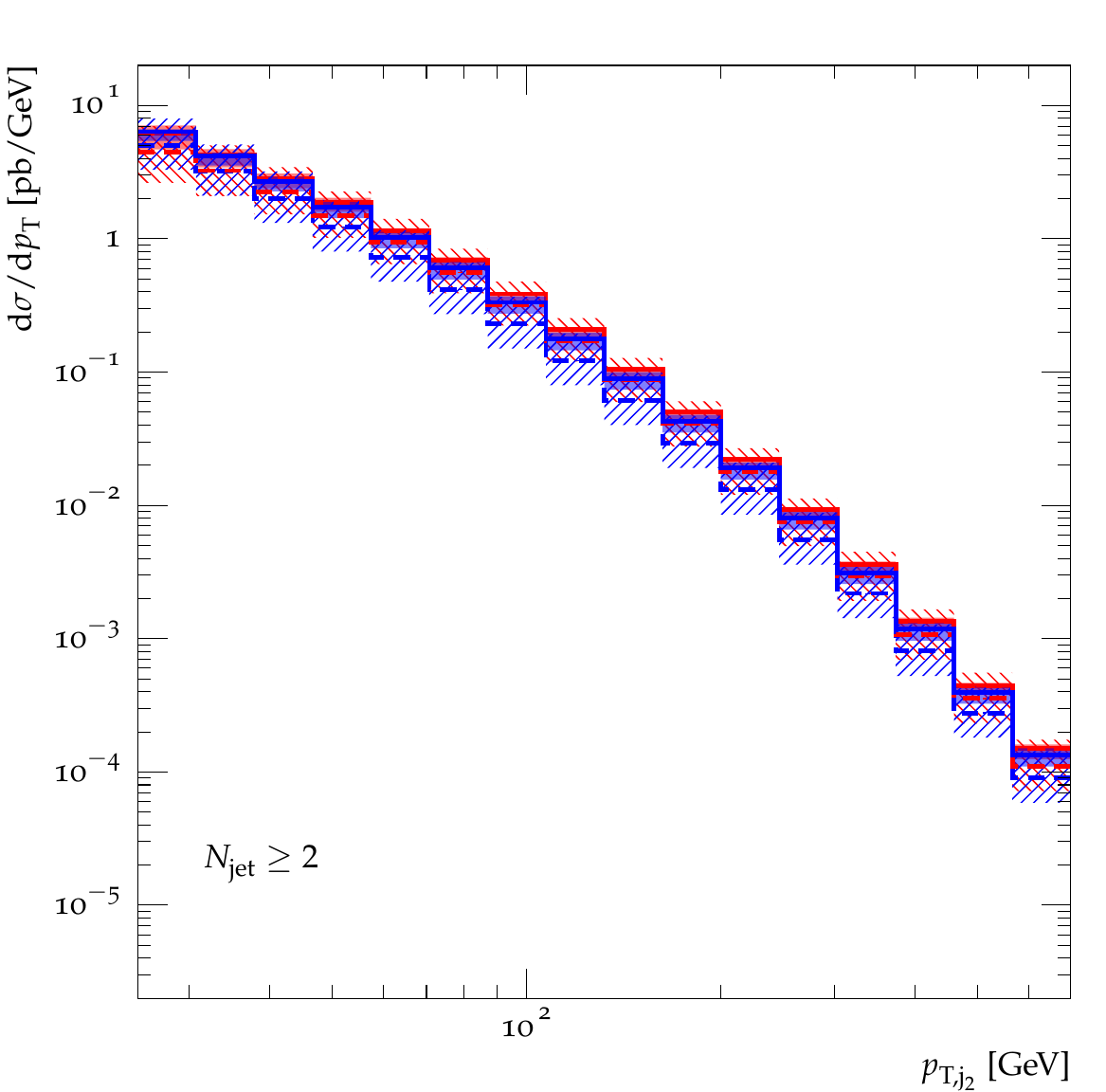}\hskipthree
    \includegraphics[scale=0.525,trim=40 25 10 0,clip]{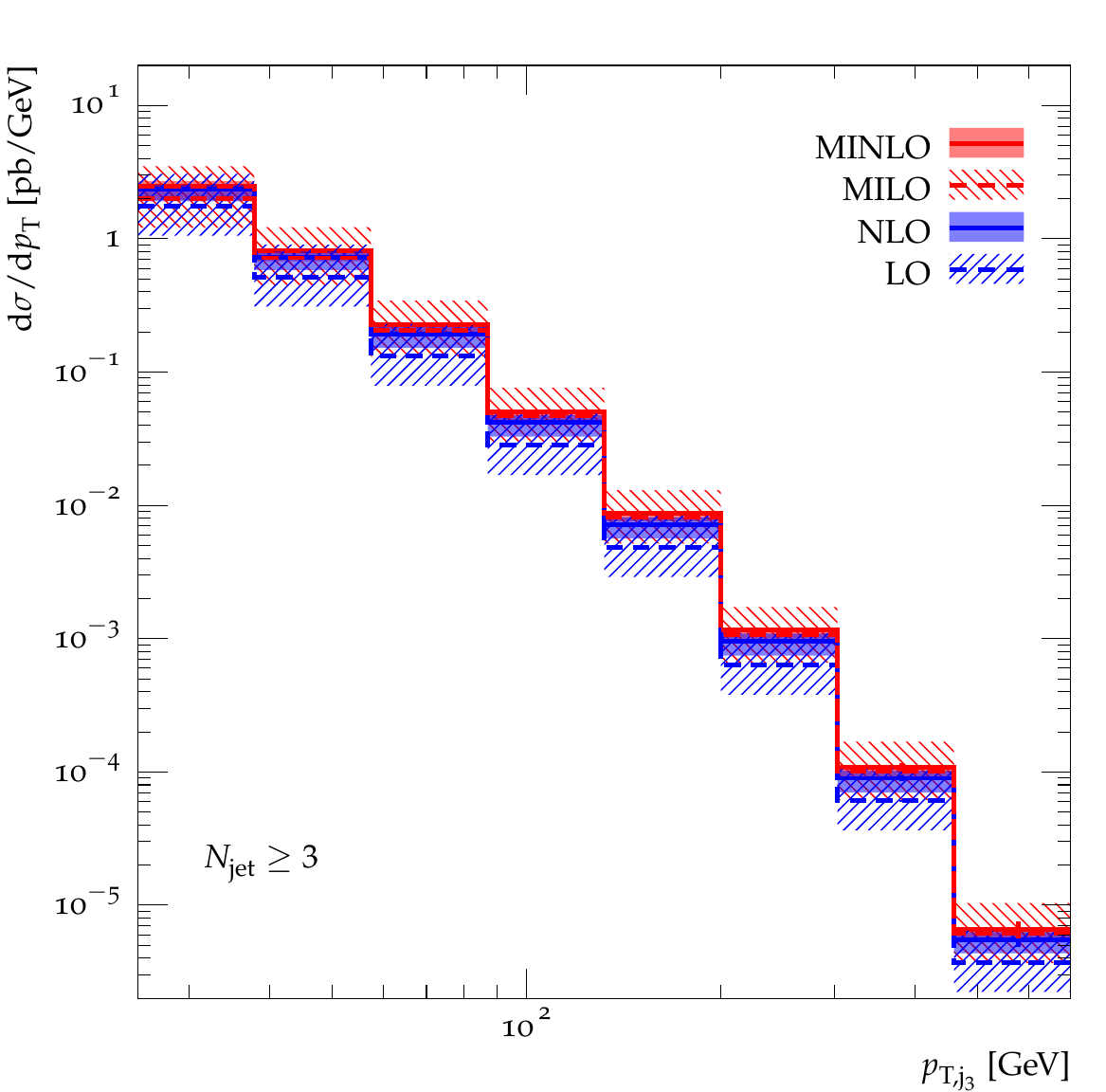}\\[-3.675pt]
    \includegraphics[scale=0.525,trim=0 25 10 0,clip]{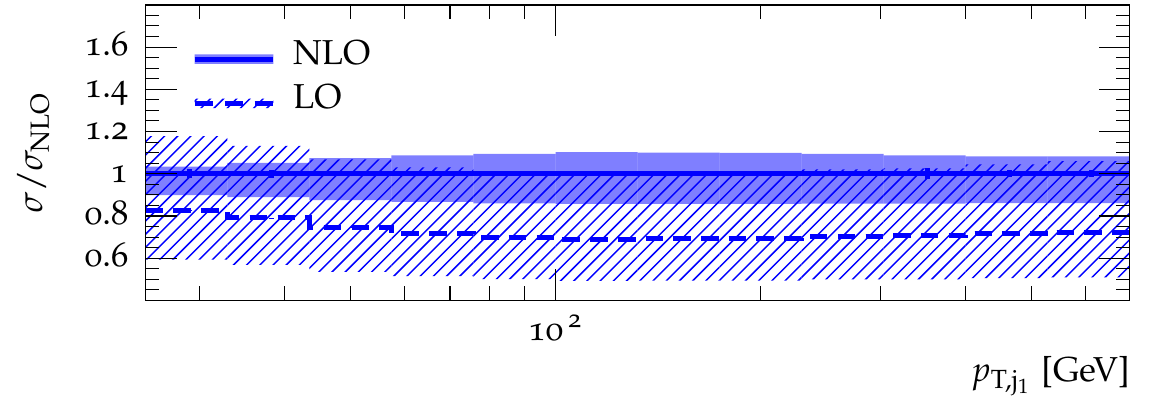}\hskipthree
    \includegraphics[scale=0.525,trim=40 25 10 0,clip]{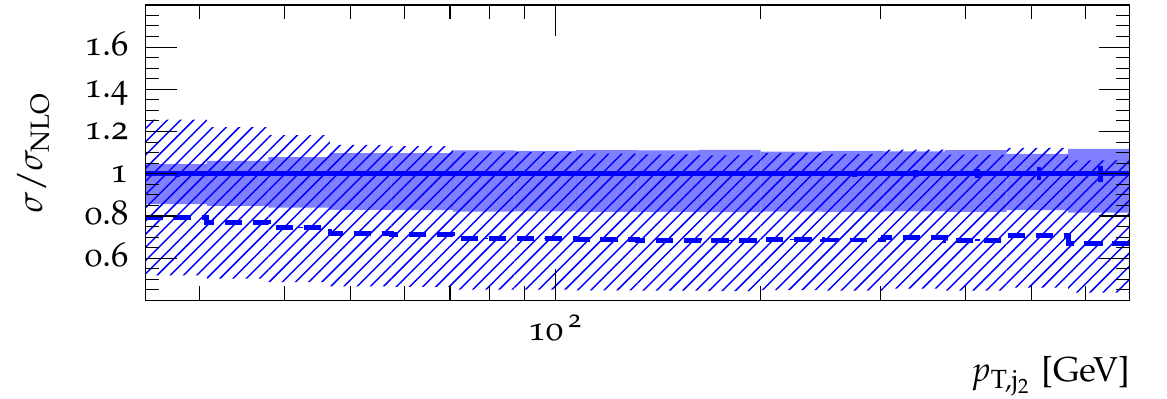}\hskipthree
    \includegraphics[scale=0.525,trim=40 25 10 0,clip]{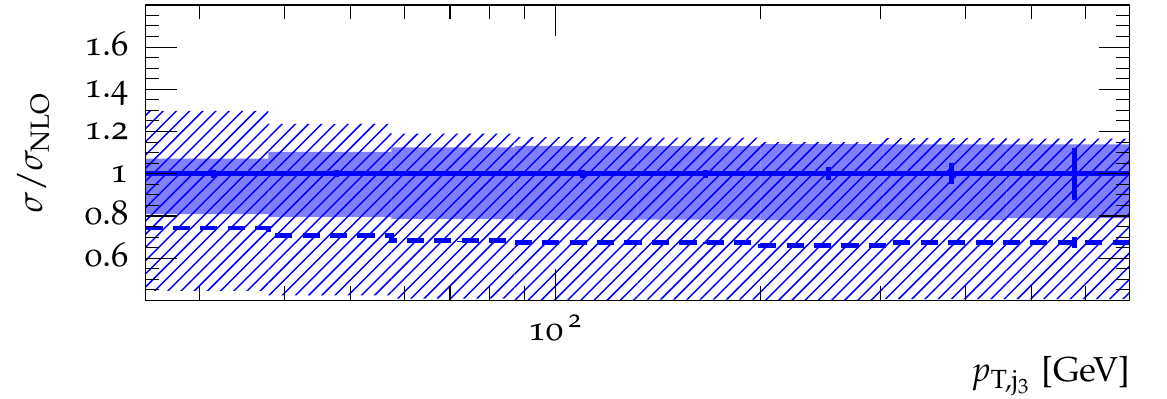}\\[-3.675pt]
    \includegraphics[scale=0.525,trim=0 25 10 0,clip]{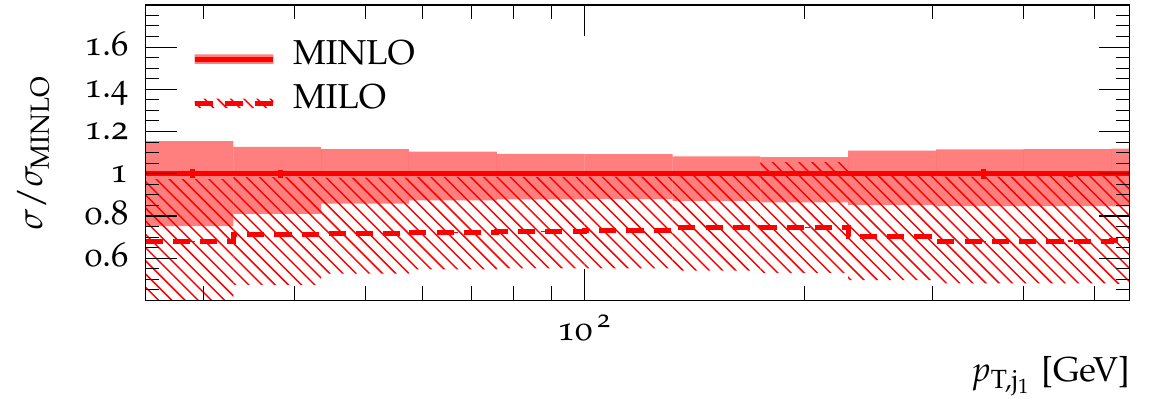}\hskipthree
    \includegraphics[scale=0.525,trim=40 25 10 0,clip]{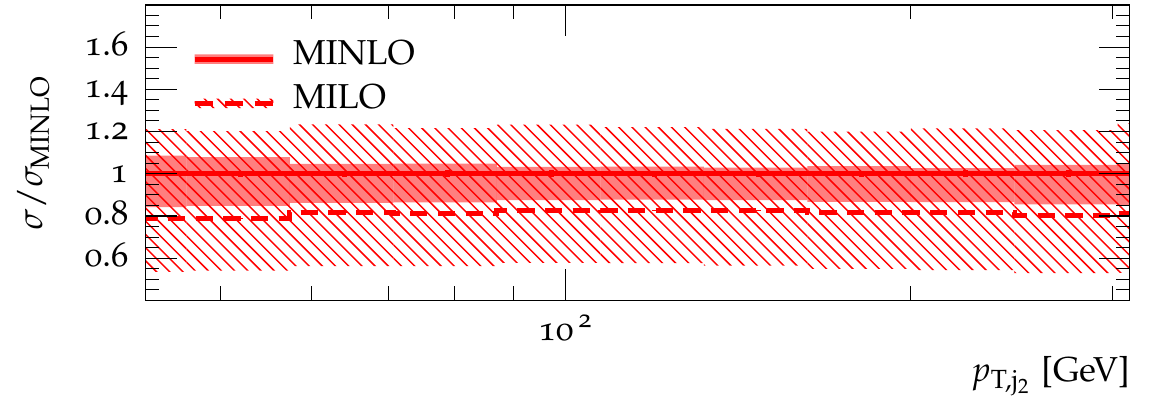}\hskipthree
    \includegraphics[scale=0.525,trim=40 25 10 0,clip]{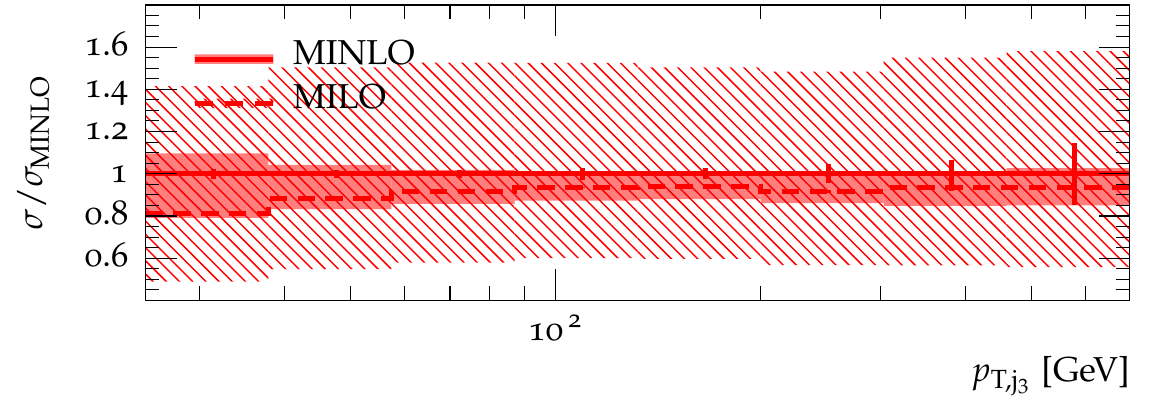}\\[-3.675pt]
    \includegraphics[scale=0.525,trim=0 0 10 0,clip]{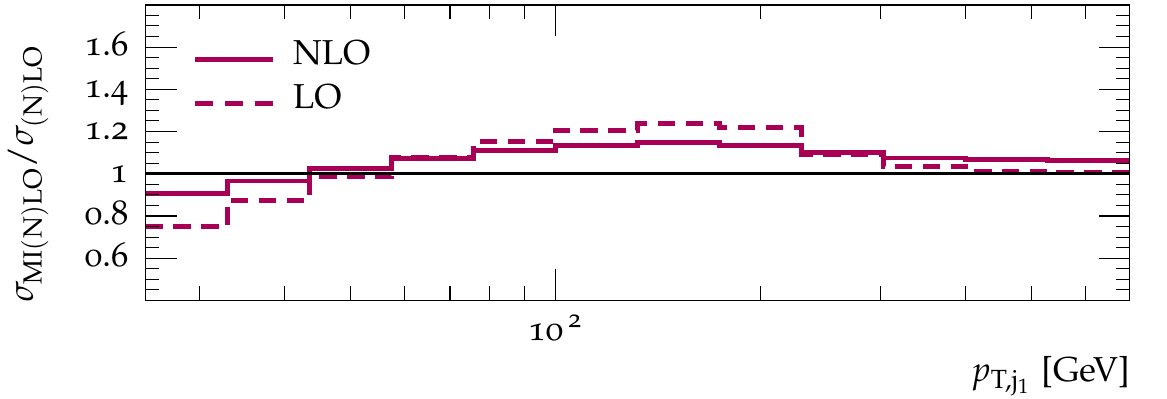}\hskipthree
    \includegraphics[scale=0.525,trim=40 0 10 0,clip]{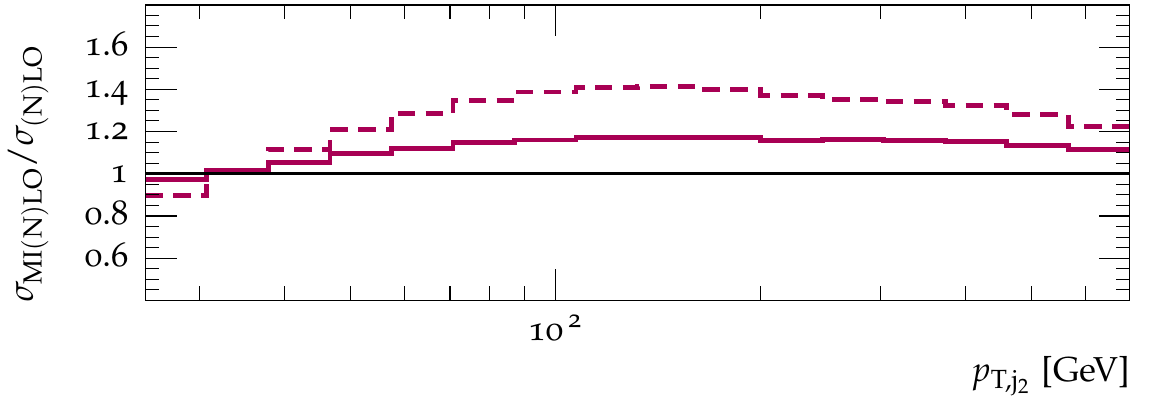}\hskipthree
    \includegraphics[scale=0.525,trim=40 0 10 0,clip]{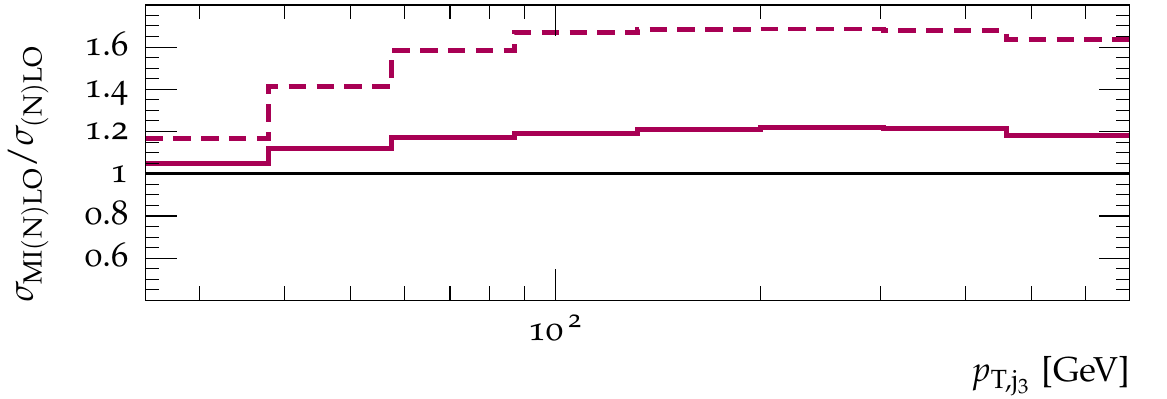}\\[3mm]
  \end{center}\capskip
  \caption{\label{fig:njet_pt}
    Distribution in the $\pT$ of the $n$-th jet for $pp\to \ttbar+n$\,jets with  $\ptjet\ge 25$\,GeV and $n=1,2,3$.}

\end{figure*}

The jet transverse momentum spectrum of the first, second and third
jet, as predicted by $\ttbar+N$\,jet calculations of corresponding jet
multiplicity, is displayed in~\reffi{fig:njet_pt}.  In general we observe
approximately constant NLO $K$-factors over the entire range of transverse
momenta analyzed here, but in terms of perturbative convergence and 
scale uncertainties at NLO we find that the \MINLO approach performs better
than fixed order.  Comparing fixed-order and \MINLO results, at LO we find
significant deviations that grow with $\njets$ and can reach 60\% in the
tails.  Such differences are largely reduced by the transition to NLO.
The fairly decent agreement between fixed-order NLO and MINLO results
exemplifies nicely how the convergence of the perturbative series leads to a
reduced dependence not only on constant scale variations, but also on the
functional form of the scale.

\begin{figure*} 
  \begin{center}
    \includegraphics[scale=0.525,trim=0 25 10 0,clip]{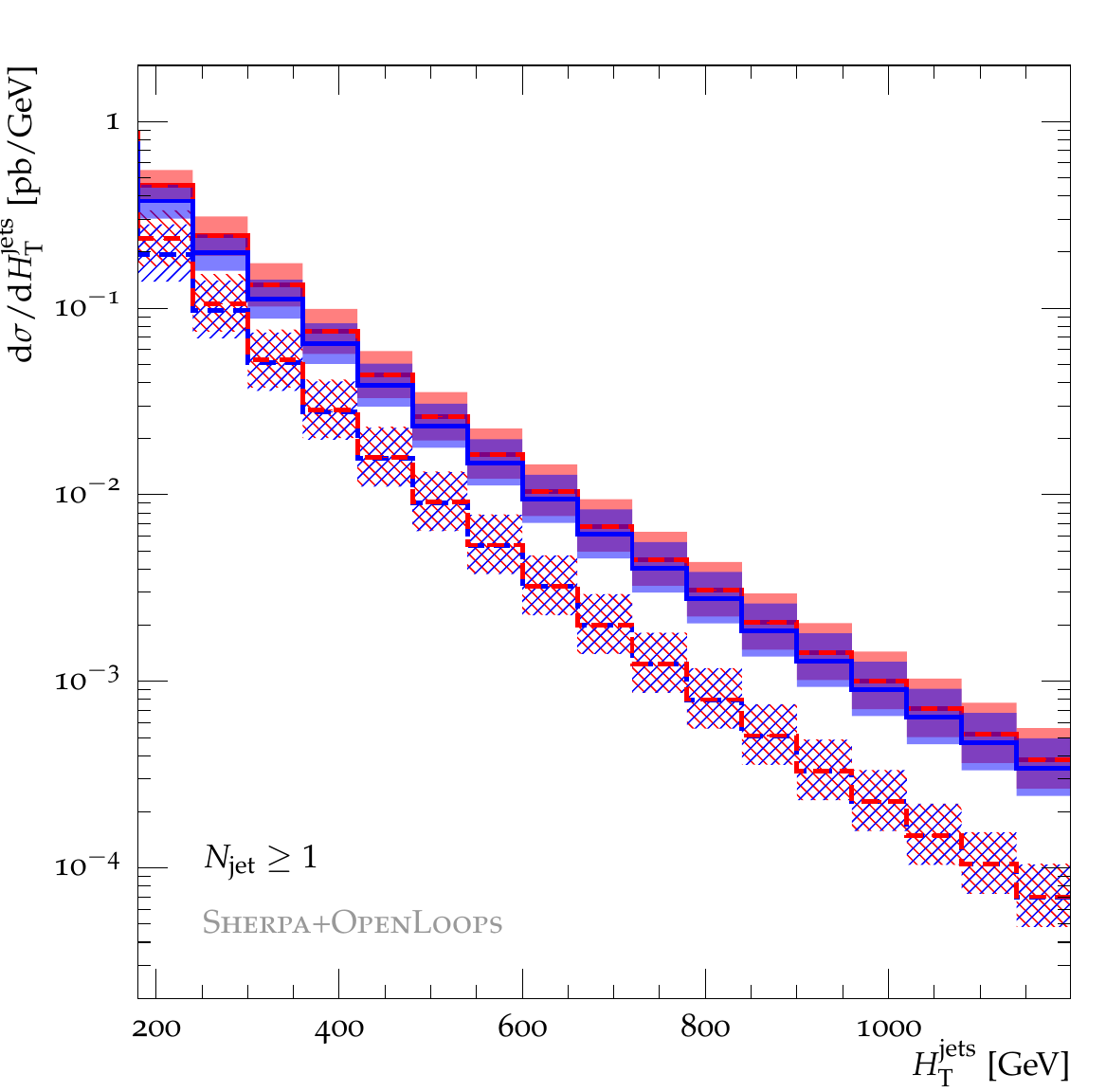}\hskipthree
    \includegraphics[scale=0.525,trim=40 25 10 0,clip]{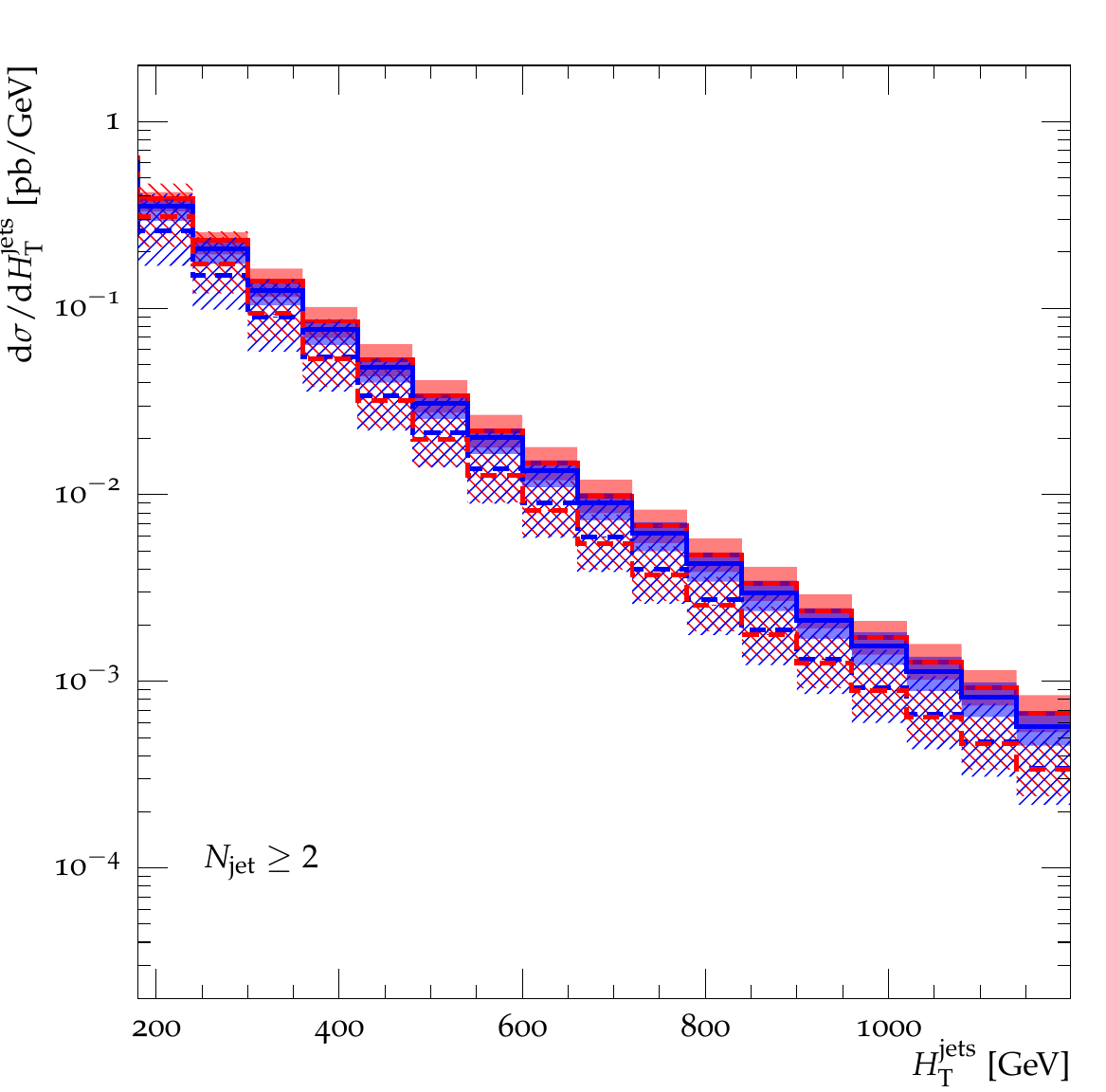}\hskipthree
    \includegraphics[scale=0.525,trim=40 25 10 0,clip]{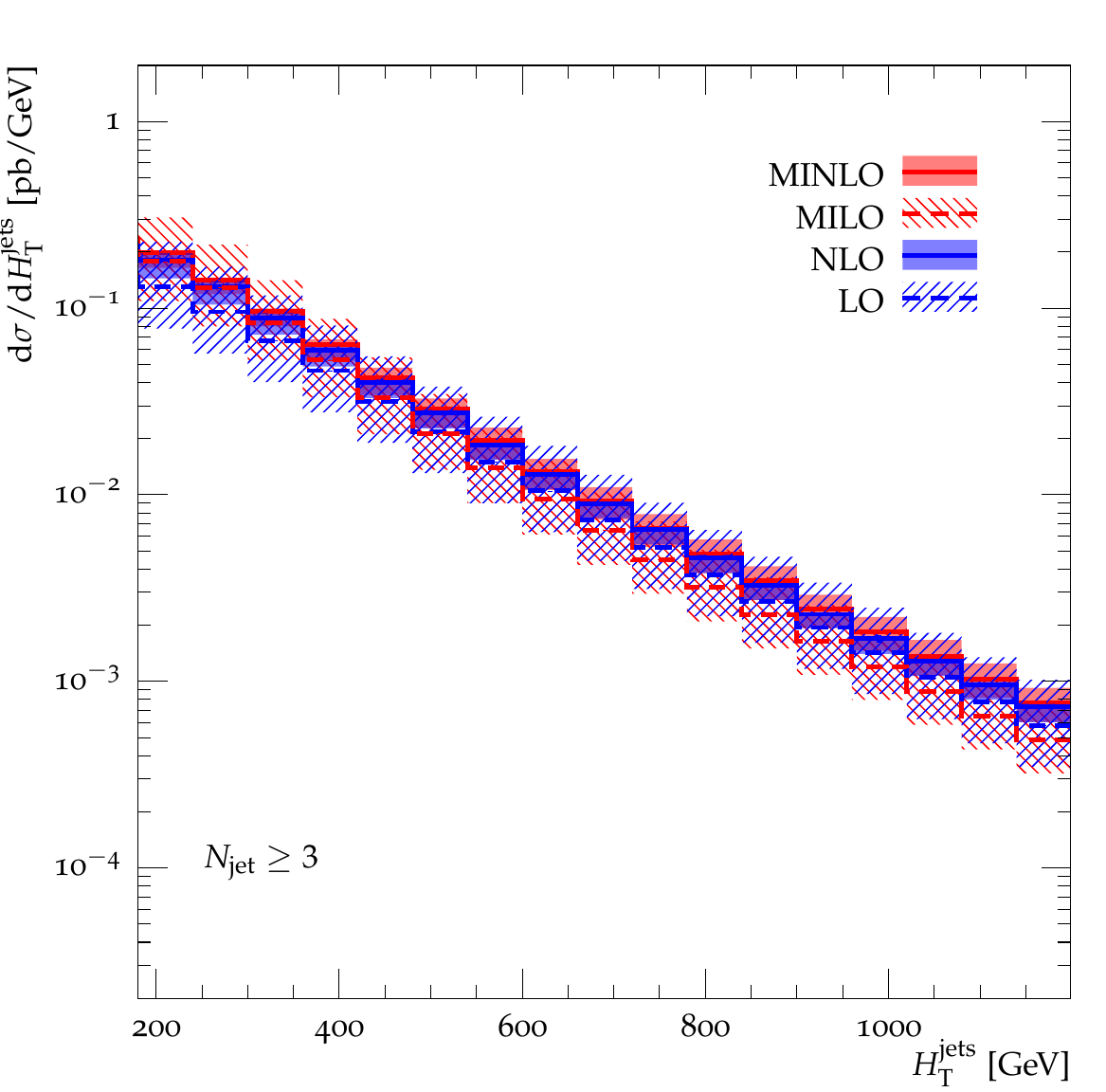}\\[-3.675pt]
    \includegraphics[scale=0.525,trim=0 25 10 0,clip]{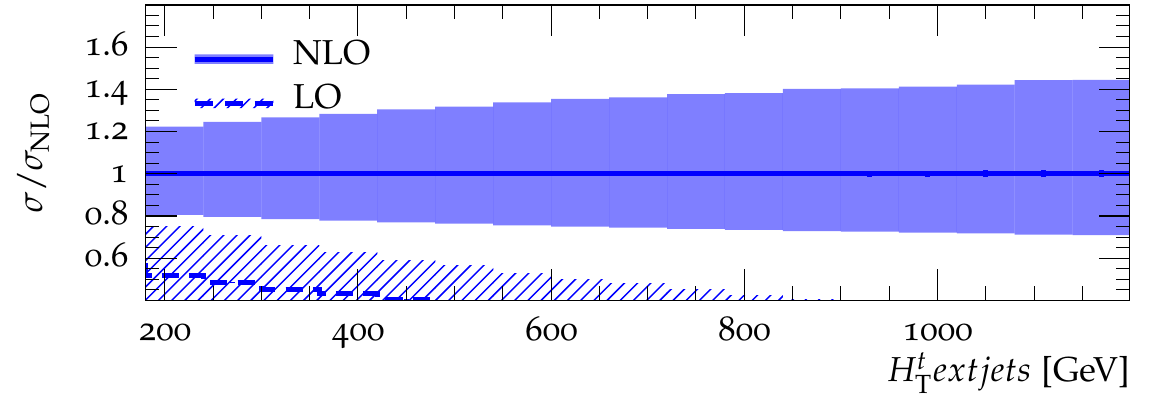}\hskipthree
    \includegraphics[scale=0.525,trim=40 25 10 0,clip]{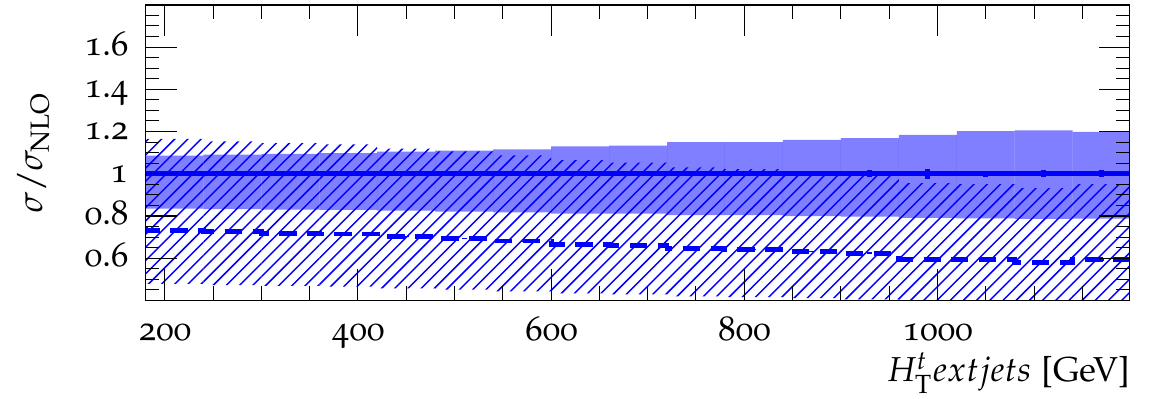}\hskipthree
    \includegraphics[scale=0.525,trim=40 25 10 0,clip]{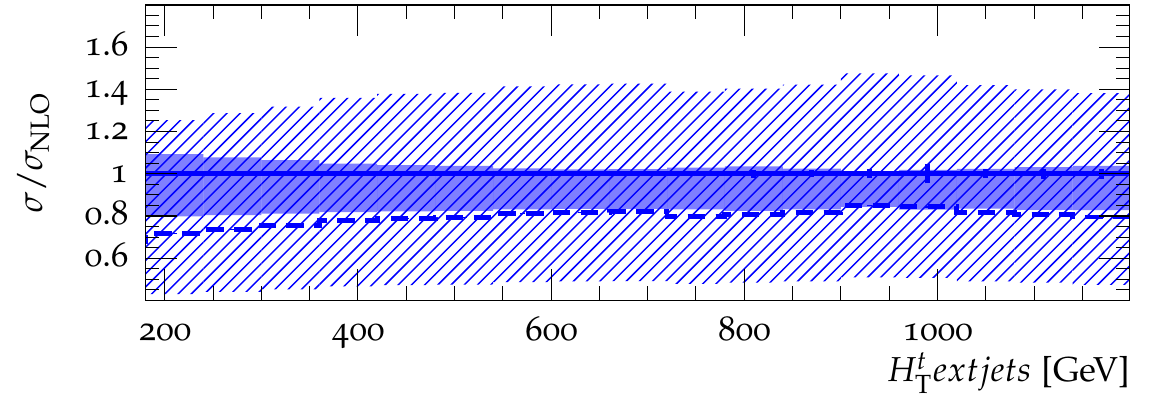}\\[-3.675pt]
    \includegraphics[scale=0.525,trim=0 25 10 0,clip]{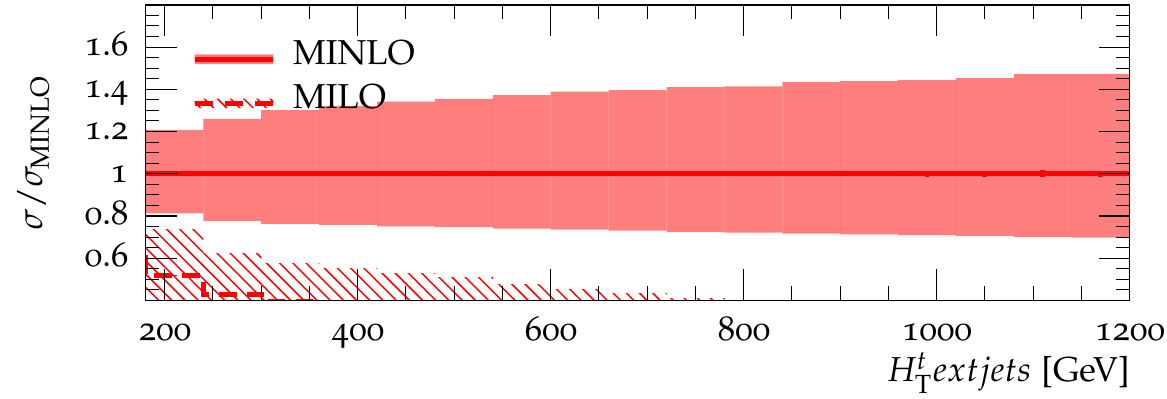}\hskipthree
    \includegraphics[scale=0.525,trim=40 25 10 0,clip]{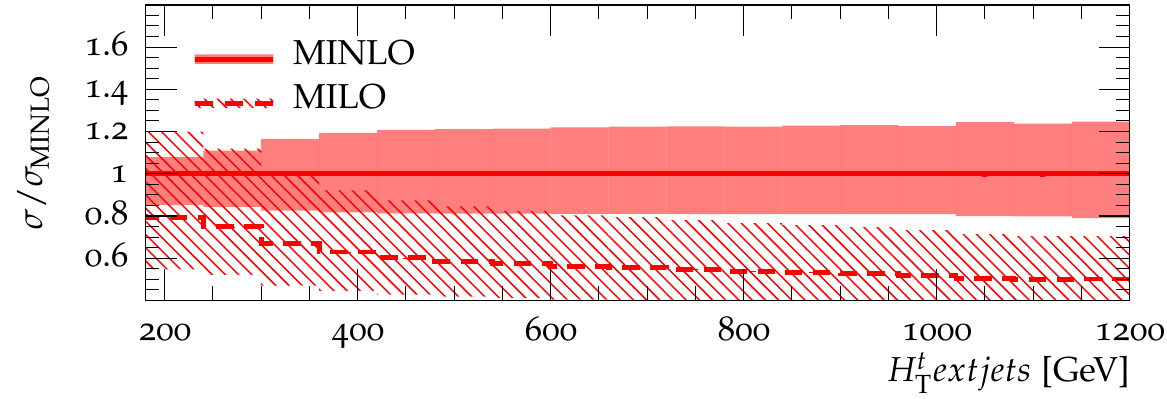}\hskipthree
    \includegraphics[scale=0.525,trim=40 25 10 0,clip]{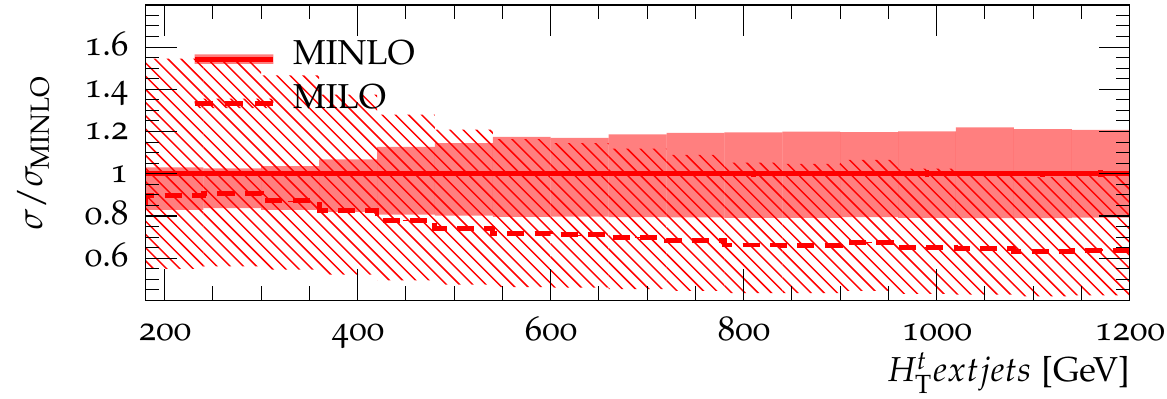}\\[-3.675pt]
    \includegraphics[scale=0.525,trim=0 0 10 0,clip]{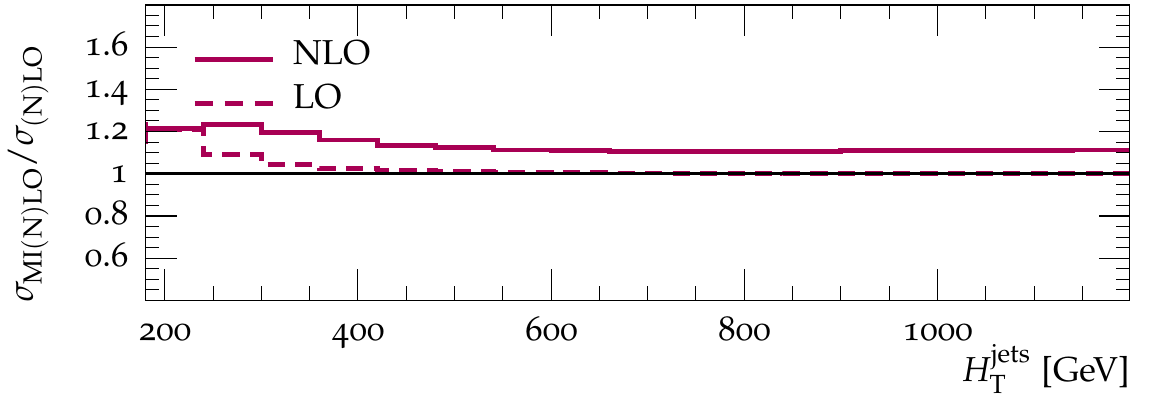}\hskipthree
    \includegraphics[scale=0.525,trim=40 0 10 0,clip]{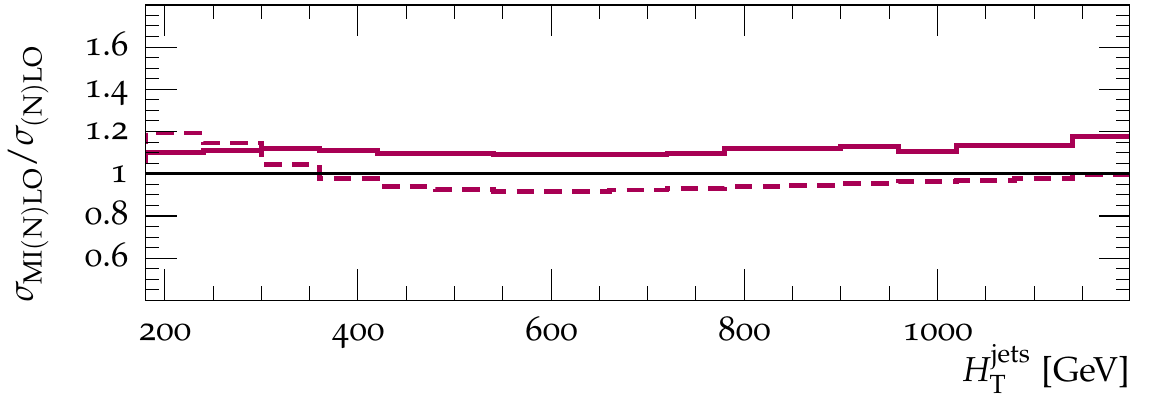}\hskipthree
    \includegraphics[scale=0.525,trim=40 0 10 0,clip]{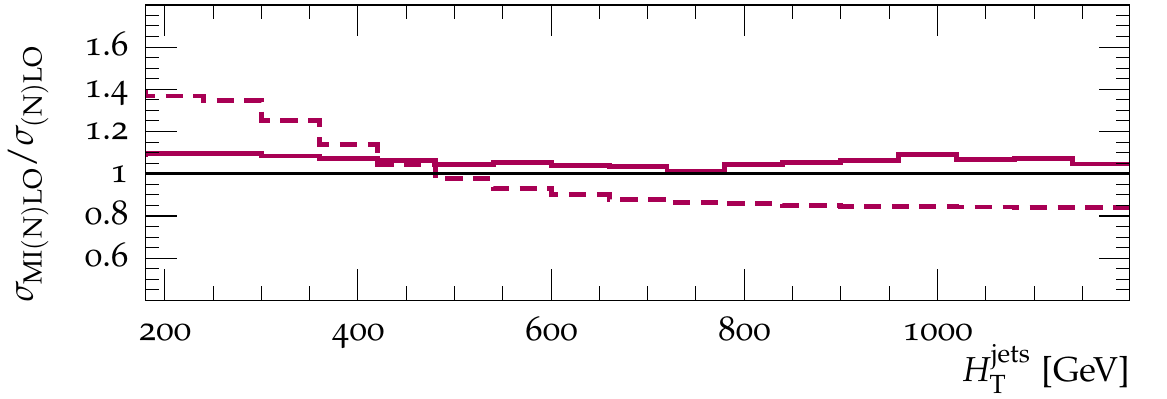}
  \end{center}\capskip
  \caption{\label{fig:njet_jetpt_ht}
Distribution in the total transverse energy of light jets for $pp\to\ttbar+1,2,3$\,jets with $\ptjet\ge 25$\,GeV.}
\end{figure*}

Figure~\ref{fig:njet_jetpt_ht} shows  inclusive 
\mbox{$\ttbar+1,2,3$\,jet} predictions for the total light-jet transverse energy,
which is defined as \mbox{$\htjets=\sum_{j} |p_{\rT,j}|$}, with the sum running over 
all reconstructed jets within acceptance.
This observable is typically badly described by LO calculations, as a sizable
fraction of events, especially at large $\htjets$, 
contains additional jets originating in initial-state 
radiation~\cite{Rubin:2010xp}. Correspondingly we observe a very large increase in the cross section 
between LO and NLO in the one-jet samples, where the effect of additional radiation not modeled by
the calculation is largest. At higher jet multiplicities, the increase is smaller, but well visible.
In \MINLO it tends to be more pronounced than at fixed order, and for $\njets\ge 3$ also MINLO 
uncertainties are larger than NLO ones.
Nevertheless, we find good overall agreement between fixed-order NLO and MINLO predictions, independent of the
jet multiplicity. 
However, given the strong sensitivity of $\htjets$ to multi-jet emissions, NLO or MINLO calculations 
with fixed jet multiplicity might significantly underestimate the effect of additional QCD radiation, 
and an approach like multijet merging at NLO~\cite{Hoeche:2014qda} would be more appropriate for this particular observable.

Studying differential distributions in several angular variables we did not
find any sizable shape effect.  We thus refrain from showing corresponding
plots.


\section{Conclusions}
We have computed predictions for top-quark pair production with up to three
additional jets at the next-to-leading order in perturbative QCD using the
automated programs \OpenLoops and \Sherpa. This is the first calculation of 
this complexity involving massive QCD partons in the final state. 
Given the multi-scale nature of $\ttbar$+multijet production, finding a scale
that guarantees optimal perturbative convergence is not trivial.
Moreover, standard factor-two scale variations might not provide a correct 
estimate of theoretical uncertainties related to missing higher-order effects.
These issues have been addressed by comparing predictions obtained at fixed order
using the scale $\HT/2$ and, alternatively, with the \MINLO method.
The hard scale $\HT/2$ is known to yield good perturbative 
convergence for a large class of processes, while the
\MINLO approach is more favorable from
the theoretical point of view, as it implements NLL resummation 
for soft and collinear logarithms that emerge 
in the presence of large ratios of scales.
For a rather wide range of observables at the 13\,TeV LHC, we find
very good agreement between the predictions generated 
at fixed order and with the \MINLO method.
The differences turn out to be well consistent with factor-two scale variations 
of the respective predictions, which are typically at the 10\% level.
These observations suggest that the fixed-order NLO and MINLO approach 
can---to a large extent---be used interchangeably.
Moreover, and most importantly, they significantly consolidate the picture of theoretical uncertainties
that results from standard scale variations alone.

\acknowledgement{
We are grateful to A.~Denner, S.~Dittmaier and L.~Hofer for providing us with pre-release
versions of the one-loop tensor-integral library \Collier.  
This research was supported by the US Department of Energy under contract
DE--AC02--76SF00515, by the Swiss National Science Foundation under
contracts BSCGI0-157722 and PP00P2-153027, by the Research Executive Agency
of the European Union under the Grant Agreements PITN--GA--2012--316704~({\it
  HiggsTools}),
by the Ka\-vli Institute for Theoretical Physics through
the National Science Foundation's Grant No. NSF PHY11-25915
and by the German Research Foundation (DFG) under grant No.\ SI 2009/1-1.
We used resources of the National Energy Research Scientific Computing
Center, which is supported by the Office of Science of the U.S.  Department
of Energy under Contract No.~DE--AC02--05CH11231.}

\bibliographystyle{amsunsrt_mod}       
\bibliography{journal}

\end{document}

%% file: table_INC_all.tex
\newcommand{\xs}[4]{${#1}(#2)^{+#3\%}_{-#4\%}$}
\def\tabspace{\hspace{3mm}}
\def\pbwidth{1cm}

\begin{tabular}{l@{\tabspace}|l@{\tabspace}c@{\tabspace}l@{\tabspace}c@{\tabspace}l@{\tabspace}l@{\tabspace}l}
$\ptjet\ge25$\,GeV                  & $\mathrm{LO}$          & $\mathrm{NLO}$        &  $\mathrm{MILO}$         & $\mathrm{MINLO}$          & \hspace{-3.5mm} $\frac{\mathrm{MINLO}}{\mathrm{NLO}}$ & \hspace{-3.5mm} $\frac{\mathrm{MINLO}}{\mathrm{MILO}}$ & \hspace{-2mm} $\frac{\mathrm{NLO}}{\mathrm{LO}}$ \\
\hline
 $\njets$\parbox{1cm}{$\ge 0$\\$=0$}
 & \xs{440.46}{22}{28}{21} & \parbox{2.5cm}{\xs{644.34}{31}{9}{11}\\\xs{375.36}{41}{3}{15}} 
 & \xs{440.46}{22}{28}{21} & \parbox{2.5cm}{\xs{683.21}{31}{12}{12}\\\xs{394.66}{41}{2}{11}} & \parbox{\pbwidth}{1.06\\1.05}  & \parbox{\pbwidth}{1.55\\0.90} & \parbox{\pbwidth}{1.46\\0.85}\\ 
 $\njets$\parbox{1cm}{$\ge 1$\\$=1$}
 & \xs{268.93}{9}{43}{28}  & \parbox{2.5cm}{\xs{361.01}{26}{7}{13}\\\xs{249.52}{27}{2}{17}}
 & \xs{267.86}{9}{36}{24}  & \parbox{2.5cm}{\xs{373.94}{23}{11}{13}\\\xs{237.88}{25}{5}{14}} & \parbox{\pbwidth}{1.04\\0.95}  & \parbox{\pbwidth}{1.40\\0.89} & \parbox{\pbwidth}{1.34\\0.93}\\ 
 $\njets$\parbox{1cm}{$\ge 2$\\$=2$}
 & \xs{111.32}{3}{59}{35}  & \parbox{2.5cm}{\xs{149.43}{16}{8}{16}\\\xs{111.11}{16}{1}{18}}
 & \xs{122.86}{4}{53}{32}  & \parbox{2.5cm}{\xs{157.49}{16}{9}{16}\\\xs{103.98}{17}{5}{21}}  & \parbox{\pbwidth}{1.05\\0.94}  & \parbox{\pbwidth}{1.28\\0.85} & \parbox{\pbwidth}{1.34\\1.00}\\ 
 $\njets$\parbox{1cm}{$\ge 3$\\$=3$}
 & \xs{38.36}{2}{75}{40}   & \parbox{2.5cm}{\xs{53.01}{10}{9}{20}\\\xs{41.12}{10}{0}{19}}
 & \xs{48.89}{2}{71}{38}   & \parbox{2.5cm}{\xs{57.43}{11}{4}{18}\\\xs{38.14}{12}{5}{38}}   & \parbox{\pbwidth}{1.08\\0.93}  & \parbox{\pbwidth}{1.17\\0.78} & \parbox{\pbwidth}{1.38\\1.07}\\ [4mm]
%
%
\hline
$\ptjet\ge40$ GeV                  & $\mathrm{LO}$          & $\mathrm{NLO}$          &  $\mathrm{MILO}$           & $\mathrm{MINLO}$           & $\frac{\mathrm{MINLO}}{\mathrm{NLO}}$ & $\frac{\mathrm{MINLO}}{\mathrm{MILO}}$ & $\frac{\mathrm{NLO}}{\mathrm{LO}}$\\
\hline
 $\njets$\parbox{1cm}{$\ge 0$\\$=0$}
 & \xs{440.46}{22}{28}{21} & \parbox{2.5cm}{\xs{644.34}{31}{9}{11}\\\xs{461.03}{36}{1}{4}}
 & \xs{440.46}{22}{28}{21} & \parbox{2.5cm}{\xs{683.21}{31}{12}{12}\\\xs{483.40}{36}{1}{5}} & \parbox{\pbwidth}{1.06\\1.05}  & \parbox{\pbwidth}{1.55 \\ 1.10} & \parbox{\pbwidth}{1.46 \\ 1.05}\\ 
 $\njets$\parbox{1cm}{$\ge 1$\\$=1$}
 & \xs{183.17}{7}{44}{28} & \parbox{2.5cm}{\xs{255.88}{20}{9}{13}\\\xs{201.57}{20}{0}{8}}
 & \xs{200.36}{7}{35}{24} & \parbox{2.5cm}{\xs{276.63}{20}{10}{12}\\\xs{206.41}{21}{3}{7}} & \parbox{\pbwidth}{1.08\\1.02}  & \parbox{\pbwidth}{1.38 \\ 1.03} & \parbox{\pbwidth}{1.40 \\ 1.10}\\ 
 $\njets$\parbox{1cm}{$\ge 2$\\$=2$}
 & \xs{54.23}{2}{59}{35} & \parbox{2.5cm}{\xs{76.13}{8}{10}{17}\\\xs{62.30}{8}{0}{12}}
 & \xs{68.34}{2}{51}{31}  & \parbox{2.5cm}{\xs{84.71}{10}{6}{14}\\\xs{63.60}{10}{2}{14}} & \parbox{\pbwidth}{1.11\\1.02}  & \parbox{\pbwidth}{1.24 \\ 0.93} & \parbox{\pbwidth}{1.40 \\ 1.15}\\ 
 $\njets$\parbox{1cm}{$\ge 3$\\$=3$}
 & \xs{13.84}{1}{75}{40} & \parbox{2.5cm}{\xs{19.87}{4}{11}{21}\\\xs{16.61}{4}{1}{16}}
 & \xs{20.55}{1}{68}{37} & \parbox{2.5cm}{\xs{22.70}{6}{2}{15}\\\xs{16.80}{6}{0}{33}} & \parbox{\pbwidth}{1.14\\1.01}  & \parbox{\pbwidth}{1.10 \\ 0.82} & \parbox{\pbwidth}{1.44 \\ 1.20}\\ [4mm]
%
%
\hline
$\ptjet\ge60$ GeV                  & $\mathrm{LO}$          & $\mathrm{NLO}$          &  $\mathrm{MILO}$           & $\mathrm{MINLO}$           & $\frac{\mathrm{MINLO}}{\mathrm{NLO}}$ & $\frac{\mathrm{MINLO}}{\mathrm{MILO}}$ & $\frac{\mathrm{NLO}}{\mathrm{LO}}$\\
\hline
 $\njets$\parbox{1cm}{$\ge 0$\\$=0$}
 & \xs{440.46}{22}{28}{21} & \parbox{2.5cm}{\xs{644.34}{31}{9}{11}\\\xs{521.32}{33}{3}{6}}
 & \xs{440.46}{22}{28}{21} & \parbox{2.5cm}{\xs{683.21}{31}{12}{12}\\\xs{547.02}{33}{5}{8}} & \parbox{\pbwidth}{1.06\\1.05}  & \parbox{\pbwidth}{1.55 \\ 1.24} & \parbox{\pbwidth}{1.46 \\ 1.18}\\ 
 $\njets$\parbox{1cm}{$\ge 1$\\$=1$}
 & \xs{123.16}{5}{44}{29} & \parbox{2.5cm}{\xs{175.69}{15}{9}{14}\\\xs{149.59}{16}{3}{10}}
 & \xs{142.02}{6}{35}{24} & \parbox{2.5cm}{\xs{195.17}{17}{9}{12}\\\xs{160.41}{17}{2}{8}} & \parbox{\pbwidth}{1.11\\1.07}  & \parbox{\pbwidth}{1.37 \\ 1.13} & \parbox{\pbwidth}{1.43 \\ 1.21}\\ 
 $\njets$\parbox{1cm}{$\ge 2$\\$=2$}
 & \xs{26.06}{1}{59}{35} & \parbox{2.5cm}{\xs{37.47}{5}{11}{18}\\\xs{32.50}{5}{4}{14}}
 & \xs{35.24}{2}{49}{30} & \parbox{2.5cm}{\xs{43.02}{8}{4}{13}\\\xs{35.10}{8}{2}{9}} & \parbox{\pbwidth}{1.15\\1.08}  & \parbox{\pbwidth}{1.22 \\ 1.00} & \parbox{\pbwidth}{1.44 \\ 1.25}\\ 
 $\njets$\parbox{1cm}{$\ge 3$\\$=3$}
 & \xs{4.95}{0}{74}{40} & \parbox{2.5cm}{\xs{7.31}{2}{13}{22}\\\xs{6.41}{2}{5}{18}}
 & \xs{7.97}{1}{65}{36} & \parbox{2.5cm}{\xs{8.61}{3}{1}{13}\\\xs{6.89}{3}{1}{27}} & \parbox{\pbwidth}{1.18\\1.07}  & \parbox{\pbwidth}{1.08 \\ 0.86} & \parbox{\pbwidth}{1.48 \\ 1.29}\\ [4mm]
%
%
\hline
$\ptjet\ge80$ GeV                  & $\mathrm{LO}$          & $\mathrm{NLO}$          &  $\mathrm{MILO}$           & $\mathrm{MINLO}$           & $\frac{\mathrm{MINLO}}{\mathrm{NLO}}$ & $\frac{\mathrm{MINLO}}{\mathrm{MILO}}$ & $\frac{\mathrm{NLO}}{\mathrm{LO}}$\\
\hline
 $\njets$\parbox{1cm}{$\ge 0$\\$=0$}
 & \xs{440.46}{22}{28}{21} & \parbox{2.5cm}{\xs{644.34}{31}{9}{11}\\\xs{555.85}{32}{5}{8}}
 & \xs{440.46}{22}{28}{21} & \parbox{2.5cm}{\xs{683.21}{31}{12}{12}\\\xs{584.21}{32}{7}{9}} & \parbox{\pbwidth}{1.06\\1.05}  & \parbox{\pbwidth} {1.55 \\ 1.33}& \parbox{\pbwidth}{1.46 \\ 1.26}\\ 
 $\njets$\parbox{1cm}{$\ge 1$\\$=1$}
 & \xs{88.46}{4}{44}{29} & \parbox{2.5cm}{\xs{127.22}{12}{10}{14}\\\xs{112.89}{12}{5}{12}}
 & \xs{104.19}{5}{34}{25} & \parbox{2.5cm}{\xs{142.99}{14}{9}{12}\\\xs{123.77}{14}{3}{9}} & \parbox{\pbwidth}{1.12\\1.10}  & \parbox{\pbwidth}{1.37 \\ 1.19}& \parbox{\pbwidth}{1.44 \\ 1.28}\\ 
 $\njets$\parbox{1cm}{$\ge 2$\\$=2$}
 & \xs{14.33}{1}{59}{35} & \parbox{2.5cm}{\xs{20.81}{3}{11}{18}\\\xs{18.64}{3}{6}{15}}
 & \xs{19.90}{1}{48}{30} & \parbox{2.5cm}{\xs{24.22}{4}{3}{12}\\\xs{20.71}{4}{1}{9}} & \parbox{\pbwidth}{1.16\\1.11}  & \parbox{\pbwidth}{1.22 \\ 1.04} & \parbox{\pbwidth}{1.45 \\ 1.30}\\ 
 $\njets$\parbox{1cm}{$\ge 3$\\$=3$}
 & \xs{2.17}{0}{74}{40} & \parbox{2.5cm}{\xs{3.22}{1}{13}{22}\\\xs{2.91}{1}{8}{19}}
 & \xs{3.59}{0}{63}{36} & \parbox{2.5cm}{\xs{3.85}{2}{1}{13}\\\xs{3.23}{2}{1}{23}} & \parbox{\pbwidth}{1.19\\1.11}  & \parbox{\pbwidth}{1.07 \\ 0.90} & \parbox{\pbwidth}{1.48 \\ 1.34}\\ 
\end{tabular}